\documentstyle[prl,aps,epsfig]{revtex}
\begin{document}
\def\be{\begin{equation}}
\def\ee{\end{equation}}
\def\d{\partial}
\def\bra{\langle}
\def\ket{\rangle}
\title{Burridge-Knopoff model and self-similarity}
\author{P.G.Akishin$^1$, M.V.Altaisky$^{1,2}$, I.Antoniou$^{3,4}$, 
A.D.Budnik$^1$, V.V.Ivanov$^{1,3}$\\ }
\address{$^1$ Laboratory of Computing Techniques and Automation, 
         Joint Institute for Nuclear Research, Dubna, 141980, RUSSIA \\
         $^2$ Space Research Institute RAS, Profsoyuznaya 84/32, 
         Moscow, 117810, RUSSIA \\
         $^3$ International Solvay Institute for Physics and Chemistry, 
         CP-231, ULB, Campus Plaine, Bd. du Triomphe, 1050, Brussels, 
         BELGIUM \\
         $^4$ Theoretische Natuurkunde, Free University of  Brussels,  
         Brussels, BELGIUM}
\maketitle
\begin{abstract}
The seismic processes are well known to be self-similar in 
both spatial and temporal behavior. At the same time, the 
Burridge-Knopoff (BK) model of earthquake fault dynamics, one of 
the basic models of theoretical seismicity,
does not posses self-similarity. In this article an extension of BK model,
which directly accounts for the self-similarity of earth crust elastic 
properties by introducing nonlinear terms for inter-block springs of BK 
model, is presented. The phase space analysis of 
the model have shown it to behave like a system of 
coupled randomly kicked oscillators. The nonlinear stiffness terms cause 
the synchronization of collective motion and produce stronger 
seismic events. 
\end{abstract}

\section{Introduction}

The earthquake -- a sudden stress relief in earth crust -- is 
believed to take place when and where the stress exceeds certain 
critical value. At the same time, it is well known that the rock 
materials are not continuous in physical sense: they consist of grains 
or crystals of different sizes separated from each other by cracks of 
all possible sizes. The picture is more or less the same from 
the typical grain sizes up to the external size of the seismic zone. 
The situation very much resembles that in developed turbulence, 
described by Kolmogorov \cite{K41a}.
Therefore, we have to deal with a system with well manifested 
self-similarity properties. The self-similarity of seismic processes 
has got a lot of attention in phenomenological studies (see e.~g. 
\cite{Kagan94,KB94} and references therein).
A theoretical account for self-similarity of earth crust fracturing 
was given by Newman and Knopoff \cite{NK90,lognorm_K} and some other authors 
on 
the base of renormalization approach. The dynamical modeling of 
self-similar velocity weakening friction in BK model was recently 
presented in \cite{SVR96}. Their hierarchical BK model accounts for 
cascade rupture propagation and seems to give a simple explanation to 
the Coulomb friction law so widely observed in nature. However, their 
model does not account for nonlinear elasticity which is also significant 
for earth crust deformations \cite{ZL95}.

The crux of our paper is the direct account of self-similarity 
(and scale-dependence, as it will be shown below) in the spirit 
of self-similar elasticity \cite{ZL95}. 
Taking self-similarity as a given property of the rock 
material, we have changed the harmonic springs $(U=\frac{kx^2}{2})$ in 
BK model \cite{BK67} to nonlinear ones possessing self-similarity 
(to be explained below). We compare our simulations with that of standard 
BK model \cite{CL89,SVR93} and found our model to be more realistic in 
the sense of foreshokes and aftershocks clustering\cite{Ger93}. 
Besides that, the nonlinear elasticity, which we have yielded from 
the self-similarity, leads to the nonlinear stiffness of the form 
$$ f(x) =-k(x)x,\quad\hbox{where}\quad 
k(x)=k_0\Bigl[ 1 +\epsilon\bigl({x\over a}+\frac{x^2}{2a^2}\bigr)
+\ldots\Bigr],$$ where $a$ is the length of relaxed spring. 
This nonlinearity is adequate to well known empirical fact of nonlinear
stiffness of the crust.

The remainder of this paper is organized as follows. In {\em section 2}
we remind the original BK model and analyze its virtues and shortcomings. 
The self-similar stiffness modification of BK model is presented in 
{\em section 3}. {\em Section 4} is devoted to the numerical algorithm 
used for simulations. In {\em section 5} we compare our model and the 
original Burridge-Knopoff one. The statistical characteristics -- 
autocorrelation function and the Hurst exponent -- are also presented 
here. In {\em section 6} we give some dynamical analysis of 
the model, which shows that the BK system behaves like a system of kicked 
oscillators.

\section{The Burridge-Knopoff model}

There is no mechanical model capable of simulating all the features of 
real seismic process. However, to study the temporal characteristics of 
earthquake dynamics is sufficient to isolate a few main 
properties of the process. The feature isolated and put as a 
basis for the mechanical model of earthquakes was a conjecture  
that earthquake faults are retarded by nonlinear friction 
between blocks of rock material. Such a purely dynamical model 
obeying the Newton laws,  without any 
{\em ad hoc} taken random forces, has been proposed 
by R.~Burridge and L.~Knopoff 30 years ago \cite{BK67}.

The BK system is a system of $N$ 
blocks of mass $m_i, i=\overline{1,N}$ rested on a rough surface and 
connected by harmonic springs of stiffness $k_c$ to each other; each 
block is attached by a leaf spring of stiffness $k_p$ to moving upper 
line, see Fig.~1. 

\vskip5mm
\begin{figure}[h]
\unitlength=1.00mm
\special{em:linewidth 0.4pt}
\linethickness{0.4pt}
\begin{picture}(144.00,46.00)
\put(10.00,8.00){\line(0,0){0.00}}
\put(10.00,8.00){\line(1,0){110.00}}
\put(20.00,10.00){\framebox(15.00,15.00)[cc]{$m_1$}}
\put(55.00,10.00){\framebox(15.00,15.00)[cc]{$m_i$}}
\put(90.00,10.00){\framebox(15.00,15.00)[cc]{$m_N$}}
\put(16.00,41.00){\rule{111.00\unitlength}{3.00\unitlength}}
\put(132.00,42.00){\vector(1,0){12.00}}
\put(45.00,22.00){\makebox(0,0)[cc]{$k_c$}}
\put(36.00,34.00){\makebox(0,0)[cc]{$k_p$}}
\put(138.00,46.00){\makebox(0,0)[cc]{$\vec v$}}
\put(32.00,25.00){\line(3,2){6.00}}
\put(38.00,29.00){\line(1,4){1.00}}
\put(39.00,33.00){\line(5,1){5.00}}
\put(44.00,34.00){\line(1,4){1.00}}
\put(45.00,38.00){\line(2,1){8.00}}
\put(100.00,25.00){\line(3,4){3.00}}
\put(103.00,29.00){\line(4,1){7.00}}
\put(110.00,30.67){\line(1,5){1.00}}
\put(111.00,36.00){\line(5,1){5.00}}
\put(116.00,37.00){\line(1,1){4.00}}
\put(35.00,17.00){\line(5,-2){5.00}}
\put(40.00,15.00){\line(3,1){9.00}}
\put(49.00,18.00){\line(1,-1){4.00}}
\put(53.00,14.00){\line(1,2){2.00}}
\put(70.00,17.00){\line(5,3){5.00}}
\put(75.00,20.00){\line(3,-4){3.00}}
\put(78.00,16.00){\line(5,3){5.00}}
\put(83.00,19.00){\line(3,-4){3.00}}
\put(86.00,15.00){\line(1,1){4.00}}
\put(66.00,25.00){\line(1,1){4.00}}
\put(70.00,29.00){\line(3,5){3.00}}
\put(73.00,34.00){\line(6,1){8.00}}
\put(81.00,35.33){\line(1,2){3.00}}
\end{picture}
\label{bk:pic}
\caption{\small The geometry of the BK model: the system is composed
of $N$ identical blocks of mass $m_i$, $k_c$ is the  stiffness of 
``horizontal'' springs,  $k_p$ is the stiffness of pulling springs, 
 $v$ is the  constant velocity of pulling line.}
\end{figure}
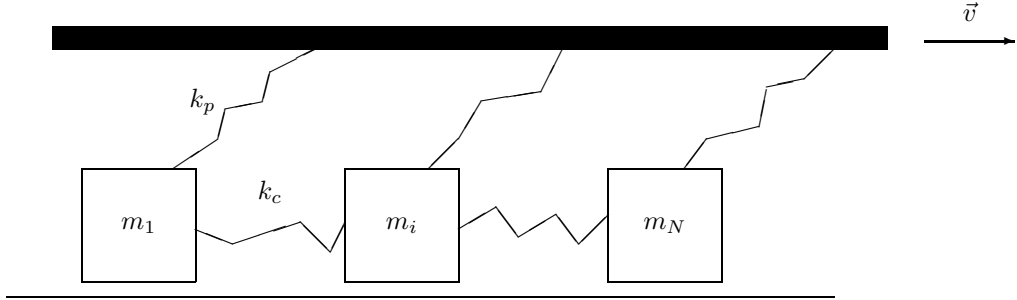

Initially 
(at $t=0$) the system is at rest, and the elastic energy accumulated 
in ``horizontal'' springs is only due to randomly generated small initial 
displacements of the blocks from their neutral positions. 
The moving upper line, which simulates the movement of external driving 
plate, exerts the force $f_n=-k_p(x_n-vt)$ on each $n$-th block. The 
nonlinear friction is defined in such a way that it holds 
each block at rest until the sum of all forces  applied to 
this block exceeds certain critical value $F_0$. Then the block makes 
a slip inhibited by nonlinear friction to a new position. A pause 
between two slips is believed to account for a pause between
earthquakes. 

Despite its simplicity, the BK model enables to simulate sequences 
of slipping events, similar to that of real earthquakes. The 
distribution of event size, generated by BK model also resembles 
the Gutenberg-Richter power law \cite{GR44}. However, for practical 
purposes of generating artificial earthquake catalogs, the ``reliability'' 
of BK toy model is not always sufficient. A number of attempts have been 
made to modify BK model to gain more realistic sequences of events, to 
gain more sharp clusterization first of all \cite{Ger93}.

Following \cite{CL89} we rewrite the equations of motion in 
using dimensionless coordinates:
\be
{\ddot{u}}_n = l^2(u_{n+1}-u_n)+l^2(u_{n-1}-u_n)-u_n+\nu\tau+F({\dot u}_n),
\label{bk}
\ee
where $u_n = \frac{k_p x_n}{F_0}$ is normalized displacement,
$l = \sqrt{k_c/k_p}$ is normalized sound velocity, 
$\nu = \frac{vk_p}{\omega_p F_0}$ is dimensionless driving 
velocity, $F_0$ is the amplitude of friction force.
The upper dots denote differentiation with respect to 
dimensionless time $\tau = \omega_p t$, $\omega_p^2 = \frac{k_p}{m}.$
The adimensional version of  friction force \cite{BK67} is taken 
in the form
\be
F({\dot u}) = \left\{ 
\begin{array}{lll}
-\frac{u}{H} & for & |\dot{u}| \le H \\
-\frac{F_0}{1+D(\dot u -H)}-E(\dot u -H) & for & \dot u > H \\
+\frac{F_0}{1-D(\dot u +H)}-E(\dot u +H) & for & \dot u < -H \\
\end{array}
\right.
\label{fric},
\ee
where $E$ is a linear friction coefficient, $H$ is a threshold velocity 
below which blocks can move without causing fracture effects; 
$D$ is a parameter of friction nonlinearity.

To keep with \cite{BK67,CL89} we have chosen $E=F_0=k_p=m=1, D=5, \nu=0.01$. 
The velocity threshold value was taken $H=0.001$.
We have performed simulations for 10, 50 and 100 blocks chains.

The BK system has several evident time scales. The characteristic time 
$T_p =\omega_p^{-1}$ is the typical time of harmonic oscillation;
$\tau_L = 1/\nu$ (dimensionless units) is the time taken by driving 
force $u-\nu\tau$  to reach the critical value $F_0=1$ and to cause a slip;
$\tau_s=N/l$ is the time required  by a ``sound'' wave traveling 
with velocity $l$ to pass over entire chain of $N$ blocks. The ratio 
of the two latter parameters $\theta = \frac{N\nu}{l}$ is an 
important parameter of the system, which accounts for the transition 
from chaotic to solitary wave regime \cite{SVR93,SVR96}.
However, in the 
discrete spring-block model (\ref{bk}) there is no dependence on 
the relaxed spring length $a$. It seems natural on one hand -- 
the seismic processes are scale-invariant practically at all scales,
-- but on the other hand the model admits the possibility of block overlapping, when 
simulated numerically. The latter is completely unphysical.

\section{Scale dependent BK model}

\subsection{Scale invariant stiffness model}

Regardless all the merits of spring-block model evidently it does not 
adequately describe the elastic properties of rock material.

Let us consider a spring of  relaxed length $a$. If we contract or
dilate it by $\Delta x$, we exhibit a reaction force $f = -k\Delta x$;
where $k$ is its stiffness. However, it is absolutely impossible to 
contract a spring up to zero length, $\Delta x =a$. More than that, as 
any geologist knows, in practice $k$ is not the same for large contractions 
and dilations. More precisely, the stiffness monotonously decreases when 
going from contractions to dilations, and then, achieving certain critical 
value, rock crashes.

The simplest way to describe such behavior is to add cubic and quartic terms 
to spring potential energy
$$
U(\Delta x)=\frac{k(\Delta x)^2}{2}+A(\Delta x)^3 + B(\Delta x)^4.
$$

However, since we are going to deal with earthquake processes, we
also have to account for such a well manifested property of the
crust as self-similarity. Following \cite{ZL95}, let us take the stiffness
to be homogeneous of degree $\epsilon$
\be
k(\lambda l)=\lambda^{-\epsilon}k(l),
\ee
normalized to the rigidity of relaxed spring $k(a)=k_0.$
The energy of the finite deformation $\Delta x$ is then given by
\be
U(\Delta x) = \int_{0}^{\Delta x} k(y)ydy = \int_{0}^{\Delta x}
\frac{a^\epsilon k_0 y dy}{(a-y)^\epsilon}.
\ee
After straightforward calculation this gives
\be
U(u) = k_0 a^2 \left[{
\frac{1}{1-\epsilon} (1-(1-u)^{1-\epsilon})
-\frac{1}{2-\epsilon} (1-(1-u)^{2-\epsilon})
}\right], \label{U}
\ee
where $u=\frac{\Delta x}{a}$ is the relative deformation.
The graph of the function $U(u)$ is presented in Fig.~2.
\vskip5mm
\begin{figure}[h]
\setlength{\unitlength}{0.240900pt}
\ifx\plotpoint\undefined\newsavebox{\plotpoint}\fi
\sbox{\plotpoint}{\rule[-0.200pt]{0.400pt}{0.400pt}}%
\begin{picture}(1500,900)(0,0)
\font\gnuplot=cmr10 at 10pt
\gnuplot
\sbox{\plotpoint}{\rule[-0.200pt]{0.400pt}{0.400pt}}%
\put(220.0,113.0){\rule[-0.200pt]{292.934pt}{0.400pt}}
\put(828.0,113.0){\rule[-0.200pt]{0.400pt}{184.048pt}}
\put(220.0,113.0){\rule[-0.200pt]{4.818pt}{0.400pt}}
\put(198,113){\makebox(0,0)[r]{0}}
\put(1416.0,113.0){\rule[-0.200pt]{4.818pt}{0.400pt}}
\put(220.0,240.0){\rule[-0.200pt]{4.818pt}{0.400pt}}
\put(198,240){\makebox(0,0)[r]{0.2}}
\put(1416.0,240.0){\rule[-0.200pt]{4.818pt}{0.400pt}}
\put(220.0,368.0){\rule[-0.200pt]{4.818pt}{0.400pt}}
\put(198,368){\makebox(0,0)[r]{0.4}}
\put(1416.0,368.0){\rule[-0.200pt]{4.818pt}{0.400pt}}
\put(220.0,495.0){\rule[-0.200pt]{4.818pt}{0.400pt}}
\put(198,495){\makebox(0,0)[r]{0.6}}
\put(1416.0,495.0){\rule[-0.200pt]{4.818pt}{0.400pt}}
\put(220.0,622.0){\rule[-0.200pt]{4.818pt}{0.400pt}}
\put(198,622){\makebox(0,0)[r]{0.8}}
\put(1416.0,622.0){\rule[-0.200pt]{4.818pt}{0.400pt}}
\put(220.0,750.0){\rule[-0.200pt]{4.818pt}{0.400pt}}
\put(198,750){\makebox(0,0)[r]{1}}
\put(1416.0,750.0){\rule[-0.200pt]{4.818pt}{0.400pt}}
\put(220.0,877.0){\rule[-0.200pt]{4.818pt}{0.400pt}}
\put(198,877){\makebox(0,0)[r]{1.2}}
\put(1416.0,877.0){\rule[-0.200pt]{4.818pt}{0.400pt}}
\put(337.0,113.0){\rule[-0.200pt]{0.400pt}{4.818pt}}
\put(337,68){\makebox(0,0){-0.8}}
\put(337.0,857.0){\rule[-0.200pt]{0.400pt}{4.818pt}}
\put(460.0,113.0){\rule[-0.200pt]{0.400pt}{4.818pt}}
\put(460,68){\makebox(0,0){-0.6}}
\put(460.0,857.0){\rule[-0.200pt]{0.400pt}{4.818pt}}
\put(582.0,113.0){\rule[-0.200pt]{0.400pt}{4.818pt}}
\put(582,68){\makebox(0,0){-0.4}}
\put(582.0,857.0){\rule[-0.200pt]{0.400pt}{4.818pt}}
\put(705.0,113.0){\rule[-0.200pt]{0.400pt}{4.818pt}}
\put(705,68){\makebox(0,0){-0.2}}
\put(705.0,857.0){\rule[-0.200pt]{0.400pt}{4.818pt}}
\put(828.0,113.0){\rule[-0.200pt]{0.400pt}{4.818pt}}
\put(828,68){\makebox(0,0){0}}
\put(828.0,857.0){\rule[-0.200pt]{0.400pt}{4.818pt}}
\put(951.0,113.0){\rule[-0.200pt]{0.400pt}{4.818pt}}
\put(951,68){\makebox(0,0){0.2}}
\put(951.0,857.0){\rule[-0.200pt]{0.400pt}{4.818pt}}
\put(1074.0,113.0){\rule[-0.200pt]{0.400pt}{4.818pt}}
\put(1074,68){\makebox(0,0){0.4}}
\put(1074.0,857.0){\rule[-0.200pt]{0.400pt}{4.818pt}}
\put(1196.0,113.0){\rule[-0.200pt]{0.400pt}{4.818pt}}
\put(1196,68){\makebox(0,0){0.6}}
\put(1196.0,857.0){\rule[-0.200pt]{0.400pt}{4.818pt}}
\put(1319.0,113.0){\rule[-0.200pt]{0.400pt}{4.818pt}}
\put(1319,68){\makebox(0,0){0.8}}
\put(1319.0,857.0){\rule[-0.200pt]{0.400pt}{4.818pt}}
\put(220.0,113.0){\rule[-0.200pt]{292.934pt}{0.400pt}}
\put(1436.0,113.0){\rule[-0.200pt]{0.400pt}{184.048pt}}
\put(220.0,877.0){\rule[-0.200pt]{292.934pt}{0.400pt}}
\put(45,495){\makebox(0,0){U(u)}}
\put(828,23){\makebox(0,0){relative deformation}}
\put(220.0,113.0){\rule[-0.200pt]{0.400pt}{184.048pt}}
\put(1306,812){\makebox(0,0)[r]{U(u)}}
\put(1328.0,812.0){\rule[-0.200pt]{15.899pt}{0.400pt}}
\put(220,357){\usebox{\plotpoint}}
\multiput(220.00,355.93)(0.669,-0.489){15}{\rule{0.633pt}{0.118pt}}
\multiput(220.00,356.17)(10.685,-9.000){2}{\rule{0.317pt}{0.400pt}}
\multiput(232.00,346.93)(0.824,-0.488){13}{\rule{0.750pt}{0.117pt}}
\multiput(232.00,347.17)(11.443,-8.000){2}{\rule{0.375pt}{0.400pt}}
\multiput(245.00,338.93)(0.669,-0.489){15}{\rule{0.633pt}{0.118pt}}
\multiput(245.00,339.17)(10.685,-9.000){2}{\rule{0.317pt}{0.400pt}}
\multiput(257.00,329.93)(0.758,-0.488){13}{\rule{0.700pt}{0.117pt}}
\multiput(257.00,330.17)(10.547,-8.000){2}{\rule{0.350pt}{0.400pt}}
\multiput(269.00,321.93)(0.669,-0.489){15}{\rule{0.633pt}{0.118pt}}
\multiput(269.00,322.17)(10.685,-9.000){2}{\rule{0.317pt}{0.400pt}}
\multiput(281.00,312.93)(0.824,-0.488){13}{\rule{0.750pt}{0.117pt}}
\multiput(281.00,313.17)(11.443,-8.000){2}{\rule{0.375pt}{0.400pt}}
\multiput(294.00,304.93)(0.758,-0.488){13}{\rule{0.700pt}{0.117pt}}
\multiput(294.00,305.17)(10.547,-8.000){2}{\rule{0.350pt}{0.400pt}}
\multiput(306.00,296.93)(0.758,-0.488){13}{\rule{0.700pt}{0.117pt}}
\multiput(306.00,297.17)(10.547,-8.000){2}{\rule{0.350pt}{0.400pt}}
\multiput(318.00,288.93)(0.824,-0.488){13}{\rule{0.750pt}{0.117pt}}
\multiput(318.00,289.17)(11.443,-8.000){2}{\rule{0.375pt}{0.400pt}}
\multiput(331.00,280.93)(0.874,-0.485){11}{\rule{0.786pt}{0.117pt}}
\multiput(331.00,281.17)(10.369,-7.000){2}{\rule{0.393pt}{0.400pt}}
\multiput(343.00,273.93)(0.758,-0.488){13}{\rule{0.700pt}{0.117pt}}
\multiput(343.00,274.17)(10.547,-8.000){2}{\rule{0.350pt}{0.400pt}}
\multiput(355.00,265.93)(0.874,-0.485){11}{\rule{0.786pt}{0.117pt}}
\multiput(355.00,266.17)(10.369,-7.000){2}{\rule{0.393pt}{0.400pt}}
\multiput(367.00,258.93)(0.950,-0.485){11}{\rule{0.843pt}{0.117pt}}
\multiput(367.00,259.17)(11.251,-7.000){2}{\rule{0.421pt}{0.400pt}}
\multiput(380.00,251.93)(0.874,-0.485){11}{\rule{0.786pt}{0.117pt}}
\multiput(380.00,252.17)(10.369,-7.000){2}{\rule{0.393pt}{0.400pt}}
\multiput(392.00,244.93)(0.874,-0.485){11}{\rule{0.786pt}{0.117pt}}
\multiput(392.00,245.17)(10.369,-7.000){2}{\rule{0.393pt}{0.400pt}}
\multiput(404.00,237.93)(0.950,-0.485){11}{\rule{0.843pt}{0.117pt}}
\multiput(404.00,238.17)(11.251,-7.000){2}{\rule{0.421pt}{0.400pt}}
\multiput(417.00,230.93)(1.033,-0.482){9}{\rule{0.900pt}{0.116pt}}
\multiput(417.00,231.17)(10.132,-6.000){2}{\rule{0.450pt}{0.400pt}}
\multiput(429.00,224.93)(0.874,-0.485){11}{\rule{0.786pt}{0.117pt}}
\multiput(429.00,225.17)(10.369,-7.000){2}{\rule{0.393pt}{0.400pt}}
\multiput(441.00,217.93)(1.033,-0.482){9}{\rule{0.900pt}{0.116pt}}
\multiput(441.00,218.17)(10.132,-6.000){2}{\rule{0.450pt}{0.400pt}}
\multiput(453.00,211.93)(1.123,-0.482){9}{\rule{0.967pt}{0.116pt}}
\multiput(453.00,212.17)(10.994,-6.000){2}{\rule{0.483pt}{0.400pt}}
\multiput(466.00,205.93)(1.033,-0.482){9}{\rule{0.900pt}{0.116pt}}
\multiput(466.00,206.17)(10.132,-6.000){2}{\rule{0.450pt}{0.400pt}}
\multiput(478.00,199.93)(1.267,-0.477){7}{\rule{1.060pt}{0.115pt}}
\multiput(478.00,200.17)(9.800,-5.000){2}{\rule{0.530pt}{0.400pt}}
\multiput(490.00,194.93)(1.123,-0.482){9}{\rule{0.967pt}{0.116pt}}
\multiput(490.00,195.17)(10.994,-6.000){2}{\rule{0.483pt}{0.400pt}}
\multiput(503.00,188.93)(1.267,-0.477){7}{\rule{1.060pt}{0.115pt}}
\multiput(503.00,189.17)(9.800,-5.000){2}{\rule{0.530pt}{0.400pt}}
\multiput(515.00,183.93)(1.267,-0.477){7}{\rule{1.060pt}{0.115pt}}
\multiput(515.00,184.17)(9.800,-5.000){2}{\rule{0.530pt}{0.400pt}}
\multiput(527.00,178.93)(1.267,-0.477){7}{\rule{1.060pt}{0.115pt}}
\multiput(527.00,179.17)(9.800,-5.000){2}{\rule{0.530pt}{0.400pt}}
\multiput(539.00,173.93)(1.378,-0.477){7}{\rule{1.140pt}{0.115pt}}
\multiput(539.00,174.17)(10.634,-5.000){2}{\rule{0.570pt}{0.400pt}}
\multiput(552.00,168.93)(1.267,-0.477){7}{\rule{1.060pt}{0.115pt}}
\multiput(552.00,169.17)(9.800,-5.000){2}{\rule{0.530pt}{0.400pt}}
\multiput(564.00,163.94)(1.651,-0.468){5}{\rule{1.300pt}{0.113pt}}
\multiput(564.00,164.17)(9.302,-4.000){2}{\rule{0.650pt}{0.400pt}}
\multiput(576.00,159.93)(1.267,-0.477){7}{\rule{1.060pt}{0.115pt}}
\multiput(576.00,160.17)(9.800,-5.000){2}{\rule{0.530pt}{0.400pt}}
\multiput(588.00,154.94)(1.797,-0.468){5}{\rule{1.400pt}{0.113pt}}
\multiput(588.00,155.17)(10.094,-4.000){2}{\rule{0.700pt}{0.400pt}}
\multiput(601.00,150.94)(1.651,-0.468){5}{\rule{1.300pt}{0.113pt}}
\multiput(601.00,151.17)(9.302,-4.000){2}{\rule{0.650pt}{0.400pt}}
\multiput(613.00,146.94)(1.651,-0.468){5}{\rule{1.300pt}{0.113pt}}
\multiput(613.00,147.17)(9.302,-4.000){2}{\rule{0.650pt}{0.400pt}}
\multiput(625.00,142.95)(2.695,-0.447){3}{\rule{1.833pt}{0.108pt}}
\multiput(625.00,143.17)(9.195,-3.000){2}{\rule{0.917pt}{0.400pt}}
\multiput(638.00,139.95)(2.472,-0.447){3}{\rule{1.700pt}{0.108pt}}
\multiput(638.00,140.17)(8.472,-3.000){2}{\rule{0.850pt}{0.400pt}}
\multiput(650.00,136.94)(1.651,-0.468){5}{\rule{1.300pt}{0.113pt}}
\multiput(650.00,137.17)(9.302,-4.000){2}{\rule{0.650pt}{0.400pt}}
\multiput(662.00,132.95)(2.472,-0.447){3}{\rule{1.700pt}{0.108pt}}
\multiput(662.00,133.17)(8.472,-3.000){2}{\rule{0.850pt}{0.400pt}}
\put(674,129.17){\rule{2.700pt}{0.400pt}}
\multiput(674.00,130.17)(7.396,-2.000){2}{\rule{1.350pt}{0.400pt}}
\multiput(687.00,127.95)(2.472,-0.447){3}{\rule{1.700pt}{0.108pt}}
\multiput(687.00,128.17)(8.472,-3.000){2}{\rule{0.850pt}{0.400pt}}
\put(699,124.17){\rule{2.500pt}{0.400pt}}
\multiput(699.00,125.17)(6.811,-2.000){2}{\rule{1.250pt}{0.400pt}}
\put(711,122.17){\rule{2.700pt}{0.400pt}}
\multiput(711.00,123.17)(7.396,-2.000){2}{\rule{1.350pt}{0.400pt}}
\put(724,120.17){\rule{2.500pt}{0.400pt}}
\multiput(724.00,121.17)(6.811,-2.000){2}{\rule{1.250pt}{0.400pt}}
\put(736,118.17){\rule{2.500pt}{0.400pt}}
\multiput(736.00,119.17)(6.811,-2.000){2}{\rule{1.250pt}{0.400pt}}
\put(748,116.67){\rule{2.891pt}{0.400pt}}
\multiput(748.00,117.17)(6.000,-1.000){2}{\rule{1.445pt}{0.400pt}}
\put(760,115.67){\rule{3.132pt}{0.400pt}}
\multiput(760.00,116.17)(6.500,-1.000){2}{\rule{1.566pt}{0.400pt}}
\put(773,114.67){\rule{2.891pt}{0.400pt}}
\multiput(773.00,115.17)(6.000,-1.000){2}{\rule{1.445pt}{0.400pt}}
\put(785,113.67){\rule{2.891pt}{0.400pt}}
\multiput(785.00,114.17)(6.000,-1.000){2}{\rule{1.445pt}{0.400pt}}
\put(797,112.67){\rule{3.132pt}{0.400pt}}
\multiput(797.00,113.17)(6.500,-1.000){2}{\rule{1.566pt}{0.400pt}}
\put(846,112.67){\rule{3.132pt}{0.400pt}}
\multiput(846.00,112.17)(6.500,1.000){2}{\rule{1.566pt}{0.400pt}}
\put(859,113.67){\rule{2.891pt}{0.400pt}}
\multiput(859.00,113.17)(6.000,1.000){2}{\rule{1.445pt}{0.400pt}}
\put(871,114.67){\rule{2.891pt}{0.400pt}}
\multiput(871.00,114.17)(6.000,1.000){2}{\rule{1.445pt}{0.400pt}}
\put(883,115.67){\rule{3.132pt}{0.400pt}}
\multiput(883.00,115.17)(6.500,1.000){2}{\rule{1.566pt}{0.400pt}}
\put(896,117.17){\rule{2.500pt}{0.400pt}}
\multiput(896.00,116.17)(6.811,2.000){2}{\rule{1.250pt}{0.400pt}}
\put(908,119.17){\rule{2.500pt}{0.400pt}}
\multiput(908.00,118.17)(6.811,2.000){2}{\rule{1.250pt}{0.400pt}}
\put(920,121.17){\rule{2.500pt}{0.400pt}}
\multiput(920.00,120.17)(6.811,2.000){2}{\rule{1.250pt}{0.400pt}}
\put(932,123.17){\rule{2.700pt}{0.400pt}}
\multiput(932.00,122.17)(7.396,2.000){2}{\rule{1.350pt}{0.400pt}}
\multiput(945.00,125.61)(2.472,0.447){3}{\rule{1.700pt}{0.108pt}}
\multiput(945.00,124.17)(8.472,3.000){2}{\rule{0.850pt}{0.400pt}}
\multiput(957.00,128.61)(2.472,0.447){3}{\rule{1.700pt}{0.108pt}}
\multiput(957.00,127.17)(8.472,3.000){2}{\rule{0.850pt}{0.400pt}}
\multiput(969.00,131.60)(1.797,0.468){5}{\rule{1.400pt}{0.113pt}}
\multiput(969.00,130.17)(10.094,4.000){2}{\rule{0.700pt}{0.400pt}}
\multiput(982.00,135.60)(1.651,0.468){5}{\rule{1.300pt}{0.113pt}}
\multiput(982.00,134.17)(9.302,4.000){2}{\rule{0.650pt}{0.400pt}}
\multiput(994.00,139.60)(1.651,0.468){5}{\rule{1.300pt}{0.113pt}}
\multiput(994.00,138.17)(9.302,4.000){2}{\rule{0.650pt}{0.400pt}}
\multiput(1006.00,143.60)(1.651,0.468){5}{\rule{1.300pt}{0.113pt}}
\multiput(1006.00,142.17)(9.302,4.000){2}{\rule{0.650pt}{0.400pt}}
\multiput(1018.00,147.59)(1.378,0.477){7}{\rule{1.140pt}{0.115pt}}
\multiput(1018.00,146.17)(10.634,5.000){2}{\rule{0.570pt}{0.400pt}}
\multiput(1031.00,152.59)(1.033,0.482){9}{\rule{0.900pt}{0.116pt}}
\multiput(1031.00,151.17)(10.132,6.000){2}{\rule{0.450pt}{0.400pt}}
\multiput(1043.00,158.59)(1.267,0.477){7}{\rule{1.060pt}{0.115pt}}
\multiput(1043.00,157.17)(9.800,5.000){2}{\rule{0.530pt}{0.400pt}}
\multiput(1055.00,163.59)(0.950,0.485){11}{\rule{0.843pt}{0.117pt}}
\multiput(1055.00,162.17)(11.251,7.000){2}{\rule{0.421pt}{0.400pt}}
\multiput(1068.00,170.59)(1.033,0.482){9}{\rule{0.900pt}{0.116pt}}
\multiput(1068.00,169.17)(10.132,6.000){2}{\rule{0.450pt}{0.400pt}}
\multiput(1080.00,176.59)(0.874,0.485){11}{\rule{0.786pt}{0.117pt}}
\multiput(1080.00,175.17)(10.369,7.000){2}{\rule{0.393pt}{0.400pt}}
\multiput(1092.00,183.59)(0.758,0.488){13}{\rule{0.700pt}{0.117pt}}
\multiput(1092.00,182.17)(10.547,8.000){2}{\rule{0.350pt}{0.400pt}}
\multiput(1104.00,191.59)(0.824,0.488){13}{\rule{0.750pt}{0.117pt}}
\multiput(1104.00,190.17)(11.443,8.000){2}{\rule{0.375pt}{0.400pt}}
\multiput(1117.00,199.59)(0.758,0.488){13}{\rule{0.700pt}{0.117pt}}
\multiput(1117.00,198.17)(10.547,8.000){2}{\rule{0.350pt}{0.400pt}}
\multiput(1129.00,207.59)(0.669,0.489){15}{\rule{0.633pt}{0.118pt}}
\multiput(1129.00,206.17)(10.685,9.000){2}{\rule{0.317pt}{0.400pt}}
\multiput(1141.00,216.58)(0.600,0.491){17}{\rule{0.580pt}{0.118pt}}
\multiput(1141.00,215.17)(10.796,10.000){2}{\rule{0.290pt}{0.400pt}}
\multiput(1153.00,226.58)(0.652,0.491){17}{\rule{0.620pt}{0.118pt}}
\multiput(1153.00,225.17)(11.713,10.000){2}{\rule{0.310pt}{0.400pt}}
\multiput(1166.00,236.58)(0.543,0.492){19}{\rule{0.536pt}{0.118pt}}
\multiput(1166.00,235.17)(10.887,11.000){2}{\rule{0.268pt}{0.400pt}}
\multiput(1178.00,247.58)(0.543,0.492){19}{\rule{0.536pt}{0.118pt}}
\multiput(1178.00,246.17)(10.887,11.000){2}{\rule{0.268pt}{0.400pt}}
\multiput(1190.00,258.58)(0.539,0.492){21}{\rule{0.533pt}{0.119pt}}
\multiput(1190.00,257.17)(11.893,12.000){2}{\rule{0.267pt}{0.400pt}}
\multiput(1203.58,270.00)(0.492,0.539){21}{\rule{0.119pt}{0.533pt}}
\multiput(1202.17,270.00)(12.000,11.893){2}{\rule{0.400pt}{0.267pt}}
\multiput(1215.58,283.00)(0.492,0.539){21}{\rule{0.119pt}{0.533pt}}
\multiput(1214.17,283.00)(12.000,11.893){2}{\rule{0.400pt}{0.267pt}}
\multiput(1227.58,296.00)(0.492,0.625){21}{\rule{0.119pt}{0.600pt}}
\multiput(1226.17,296.00)(12.000,13.755){2}{\rule{0.400pt}{0.300pt}}
\multiput(1239.58,311.00)(0.493,0.576){23}{\rule{0.119pt}{0.562pt}}
\multiput(1238.17,311.00)(13.000,13.834){2}{\rule{0.400pt}{0.281pt}}
\multiput(1252.58,326.00)(0.492,0.669){21}{\rule{0.119pt}{0.633pt}}
\multiput(1251.17,326.00)(12.000,14.685){2}{\rule{0.400pt}{0.317pt}}
\multiput(1264.58,342.00)(0.492,0.755){21}{\rule{0.119pt}{0.700pt}}
\multiput(1263.17,342.00)(12.000,16.547){2}{\rule{0.400pt}{0.350pt}}
\multiput(1276.58,360.00)(0.493,0.695){23}{\rule{0.119pt}{0.654pt}}
\multiput(1275.17,360.00)(13.000,16.643){2}{\rule{0.400pt}{0.327pt}}
\multiput(1289.58,378.00)(0.492,0.841){21}{\rule{0.119pt}{0.767pt}}
\multiput(1288.17,378.00)(12.000,18.409){2}{\rule{0.400pt}{0.383pt}}
\multiput(1301.58,398.00)(0.492,0.884){21}{\rule{0.119pt}{0.800pt}}
\multiput(1300.17,398.00)(12.000,19.340){2}{\rule{0.400pt}{0.400pt}}
\multiput(1313.58,419.00)(0.492,0.970){21}{\rule{0.119pt}{0.867pt}}
\multiput(1312.17,419.00)(12.000,21.201){2}{\rule{0.400pt}{0.433pt}}
\multiput(1325.58,442.00)(0.493,0.972){23}{\rule{0.119pt}{0.869pt}}
\multiput(1324.17,442.00)(13.000,23.196){2}{\rule{0.400pt}{0.435pt}}
\multiput(1338.58,467.00)(0.492,1.099){21}{\rule{0.119pt}{0.967pt}}
\multiput(1337.17,467.00)(12.000,23.994){2}{\rule{0.400pt}{0.483pt}}
\multiput(1350.58,493.00)(0.492,1.272){21}{\rule{0.119pt}{1.100pt}}
\multiput(1349.17,493.00)(12.000,27.717){2}{\rule{0.400pt}{0.550pt}}
\multiput(1362.58,523.00)(0.493,1.250){23}{\rule{0.119pt}{1.085pt}}
\multiput(1361.17,523.00)(13.000,29.749){2}{\rule{0.400pt}{0.542pt}}
\multiput(1375.58,555.00)(0.492,1.530){21}{\rule{0.119pt}{1.300pt}}
\multiput(1374.17,555.00)(12.000,33.302){2}{\rule{0.400pt}{0.650pt}}
\multiput(1387.58,591.00)(0.492,1.789){21}{\rule{0.119pt}{1.500pt}}
\multiput(1386.17,591.00)(12.000,38.887){2}{\rule{0.400pt}{0.750pt}}
\multiput(1399.58,633.00)(0.492,2.090){21}{\rule{0.119pt}{1.733pt}}
\multiput(1398.17,633.00)(12.000,45.402){2}{\rule{0.400pt}{0.867pt}}
\multiput(1411.58,682.00)(0.493,2.439){23}{\rule{0.119pt}{2.008pt}}
\multiput(1410.17,682.00)(13.000,57.833){2}{\rule{0.400pt}{1.004pt}}
\multiput(1424.58,744.00)(0.492,3.900){21}{\rule{0.119pt}{3.133pt}}
\multiput(1423.17,744.00)(12.000,84.497){2}{\rule{0.400pt}{1.567pt}}
\put(810.0,113.0){\rule[-0.200pt]{8.672pt}{0.400pt}}
\end{picture}
\label{scpot:pic}
\vskip5mm
\caption{\small The graph of the potential energy of a 
spring with self-similar stiffness 
$k(\lambda a) = \lambda^{-\epsilon}k(a)$, function (\ref{U}). 
$X$-axis values correspond to relative deformation $u=\frac{\Delta x}{a}$,
where $a$ is the length of relaxed spring. Plotted 
for $\epsilon=0.5, k_0=a=1$.}
\end{figure}
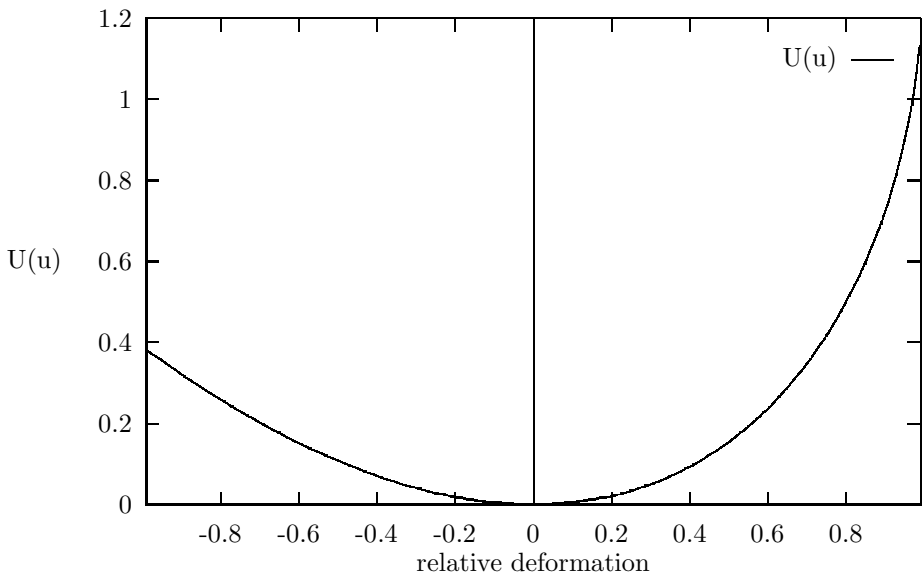

We will proceed with a case of small nonlinearity, which is
the restriction of general formula (\ref{U}) for the case of small
$\epsilon$, and has convenient linear stiffness $k=const$
as a limiting case at $\epsilon\to0$. We will concentrate ourselves 
on elastic nonlinearity without taking friction into account at this 
point. Taking the Taylor expansion of (\ref{U}) with respect 
to $\epsilon$
\be
U(u) = k_0a^2 \Bigl[{
\frac{u^2}{2}
+ \epsilon\bigl(
{\frac{u^3}{3}+\frac{u^4}{8}+\frac{u^5}{15}}
\bigr) 
+ \epsilon^2\bigl(
{\frac{u^4}{8}+\frac{u^5}{10}}
\bigr) +  o(\epsilon^3)}\Bigr].
\label{U2}
\ee
In the first order with respect to $\epsilon$ we get
$f(x) = -\frac{\d U}{\d x} = -k_1(x)x,$ where
$$
k_1(x)=k_0\Bigl[1+\epsilon\bigl({x\over a}+\frac{x^2}{2a^2}
+\frac{x^4}{3a^3}\bigr) \Bigr]
$$
is the deformation dependent spring strength. Therefore, the self-similarity 
of elastic media necessarily causes nonlinearity.

Using the nonlinear potential (\ref{U2}) for small displacements ($u\ll 1$)
we have 
\be
U(u) = k_0a^2 \Bigl[{
\frac{u^2}{2}
+ \epsilon\bigl({\frac{u^3}{3}+\frac{u^4}{8}}\bigr) 
+  o(\epsilon^2)}\Bigr],
\label{U1}
\ee
as a spring potential, we account for small influence of self-similar 
stiffness ($\epsilon\ll1$). 

The first term of the equation (\ref{U1}), the harmonic spring potential, is already 
contained in the BK model. The nonlinearity comes from the second term. The 
force applied to the $n$-th block of the spring-block chain due to this 
nonlinear term is  
\be
f_{nlin}(x_n) = -\frac{\d}{\d x_n} U_{nlin}, \quad 
 U_{nlin} = \sum_n \frac{k_c\epsilon}{3a}\Delta_n^3 
                 +\frac{k_c\epsilon}{8a^2}\Delta_n^4,
\label{nlf} 
\ee
where $\Delta_n=x_n-x_{n+1}$.
So, the nonlinear stiffness term will appear at the r.~h.~s. 
of equation of motion (\ref{bk}): 
$$
m\ddot{x}_n = k_c(x_{n+1}-2x_n + x_{n-1}) -k_p(x_n-vt)
+f_{nlin}(x_n)+\hat F(\dot{x_n}),
$$
where $f_{nlin}(x_n)$, given by (\ref{nlf}). In dimensionless form the 
spring-block equation (\ref{bk}) with nonlinear terms described above 
takes the form 
\begin{eqnarray}
\nonumber
\ddot{u}_n &=& l^2(u_{n+1}-u_n) - l^2(u_n-u_{n-1})) -u_n +\nu\tau \\
           &+& \alpha\bigl[(u_{n+1}-u_n)^2-(u_n-u_{n-1})^2\bigr] \\
\nonumber  &+& \beta \bigl[(u_{n+1}-u_n)^3-(u_n-u_{n-1})^3\bigr] 
+F(\dot{u}_n),
\label{bknldl}
\end{eqnarray}
where the dimensionless nonlinearity parameters are
$$\alpha 
         = -\frac{l^2\epsilon F_0}{a k_p}, \quad
  \beta  
         = \frac{l^2\epsilon F_0^2}{2a^2 k_p^2};$$
all other notations are the same as for the equation (\ref{bk}).

\section{Numerical implementation}
In this section we consider the Burridge-Knopoff model consisting on $N$ 
blocks. For numerical simulations we rewrite equation (\ref{bk}) separately 
for 
boundary  ($i=1,N$) and internal blocks ($1<i<N$) motion 
in the following form:
\be 
\begin{array}{lcl}
\ddot{x_1} &=& l^2 ( x_{2} - x_{1}) - ( x_{1} - \nu t ) + \Psi_1(x_1,x_2, {\dot x}_1), \\
{\ddot{x}}_N &=& l^2 ( x_{N-1} - x_{N}) - ( x_{N} - \nu t ) + \Psi_N(x_{N},x_{N-1}, {\dot x}_N),\\
\ddot{x_i} &=& l^2 ( x_{i-1} - 2x_{i} + x_{i-1} ) - ( x_{i} - \nu t )+ \Psi_i(x_{i-1},x_i,x_{i+1}, {\dot x}_i),\\
           & &\quad i=2,\dots,N-1,  
\end{array}
\label{bkns}
\ee
where functions $\Psi_i$ comprise nonlinear terms of r.~h.~s. of (\ref{bk}).
Free boundary conditions were set for the first and the last blocks.

The problem is reduced to the solution of the nonlinear equations of 
motion (\ref{bkns}) with given initial conditions:
\be x_i(t_0)= u_i^{\>0},
\qquad
{\dot x}_i(t_0)= v_i^{\>0}, \quad i=1,\dots,N.
\label{bc1}
\ee

Let $p_i(t) = {\dot x}_i(t)$. We rewrite the system (\ref{bkns},\ref{bc1}) 
in the following form:
\begin{eqnarray}
\nonumber        {\dot x}_i &=& p_{i}(t), \quad i=1,\dots,N, \\
\nonumber        {\dot p}_1 &=& l^2 ( x_{2} - x_{1}) - ( x_{1} - \nu t ) + \Psi_1(x_1,x_2, {p}_1), \\
\label{cauchy}   {\dot p}_N &=& l^2 ( x_{N-1} - x_{N}) - ( x_{N} - \nu t ) + \Psi_N(x_N,x_{N-1}, {p}_N), \\
\nonumber        {\dot p}_i &=& l^2( x_{i-1} - 2x_{i} + x_{i-1} ) - ( x_{i} - \nu t )
                             + \Psi_i(x_{i-1},x_i,x_{i+1}, {p}_i), \\
\nonumber                            & &\quad i=2,\dots,N-1,  
\end{eqnarray}
with initial conditions:
\be x_i(t_0)= u_i^{\>0},
\qquad
 p_i(t_0)= v_i^{\>0}, \quad i=1,\dots,N
\label{ic}.
\ee

This scheme has been used for numerical simulations of both linear BK 
model ($\epsilon=0$) and that with nonlinear terms  ($\epsilon >0$). 
The nonlinear friction was given by (\ref{fric}). 
The initial displacements $u_i$ were randomly generated with the amplitude 
0.0001, the initial velocities were set to zero, we choose $l^2=100$.
\def\dt{\delta t}
Let $\dt$ be the time step $t_{l}=t_{0}+l\dt$.
To get an approximate solution of the Cauchy problem
(\ref{cauchy}, \ref{ic})
we use  the implicit numerical scheme:
\begin{eqnarray}
\nonumber \frac{x_{i}^{l}- x_{i}^{l-1}}{\dt}  &=& 0.5(p_{i}^{l}+p_{i}^{l-1}), \quad i=1,\dots,N, \\ 
\nonumber \frac{({p_1}^{l}-{p_1}^{l-1})}{\dt} &=& 0.5(l^2(x_{2}^{l}-x_{1}^{l})- x_{1}^{l}+ 
                                                  l^2(x_{2}^{l-1}-x_{1}^{l-1})- x_{1}^{l-1)})\\            
\nonumber  &+&\nu t  +  \Psi_1(x_1^{l-1},x_2^{l-1}, {p}_1^{l-1}),\\ 
\nonumber \frac{({p_N}^{l}-{p_N}^{l-1})}{\dt} &=& 0.5(l^2(x_{N-1}^{l}-x_{N}^{l})- x_{N}^{l}+ 
                                                  l^2(x_{N-1}^{l-1}-x_{N}^{l-1})- x_{N}^{l-1)})\\            
            &+&\nu t  +  \Psi_N(x_N^{l-1},x_{N-1}^{l-1}, {p}_N^{l-1}), \label{scheme} \\ 
\nonumber  \frac{({p_i}^{l}-{p_i}^{l-1})}{\dt} &=&0.5(l^2(x_{i-1}^{l}-2x_{i}^{l}+x_{i+1}^{l})
                  - x_{i}^{l}+ l^2(x_{i-1}^{l-1}-2x_{i}^{l-1}+x_{i+1}^{l-1})-x_{i}^{l-1})\\ 
\nonumber  &+&\nu t + \Psi_i(x^{l-1}_{i-1},x^{l-1}_i,x^{l-1}_{i+1}, {p}^{l-1}_i), 
\quad i=2,\dots,N-1,  
\end{eqnarray}
with initial conditions (\ref{ic}).  
To keep with \cite{BK67,CL89} we have chosen
$E=F_0=k_p=m=1, D=5, \nu=0.01$. 
The velocity threshold value was taken $H=0.001$.

\section{Results}

\subsection{General outlook}

The dynamical behavior of both linear (\ref{bk}) and nonlinear 
(\ref{bknldl}) models was studied in terms of potential and 
kinetic energy. We calculated the time dependence of potential $U$ and kinetic 
$T$ energies:
\begin{eqnarray} 
\nonumber U_c(\tau) &=& {l^2\over2} \sum_{i=1}^{N-1} (u_{i+1}(\tau)-u_{i}(\tau))^2, \\
\nonumber &-& \frac{l^2\epsilon}{3}\sum_{i=1}^{N-1}(u_{i+1}(\tau)-u_{i}(\tau))^3+ \frac{l^2\epsilon}{8}\sum_{i=1}^{N-1}(u_{i+1}(\tau)-u_{i}(\tau))^4,\\ 
\label{pe} U_m(\tau) &=&  {1\over2} \sum_{i=1}^{N}(\nu\tau-u_{i}(\tau))^2, \\
\nonumber U_{pot} &=& U_c + U_m, \\
\label{ke} T(\tau) &=& {1\over2} \sum_{i=1}^{N}{\dot{u}}_{i}^2(\tau).
\end{eqnarray}

The dependence of the energies  on dimensionless time
 $\tau$ is shown in Figs.~3 and 4, for linear ($\epsilon =0$) and 
nonlinear ($\epsilon =0.2$) model, respectively.
\begin{figure}[ht]
\centerline{\mbox{\epsfig{file=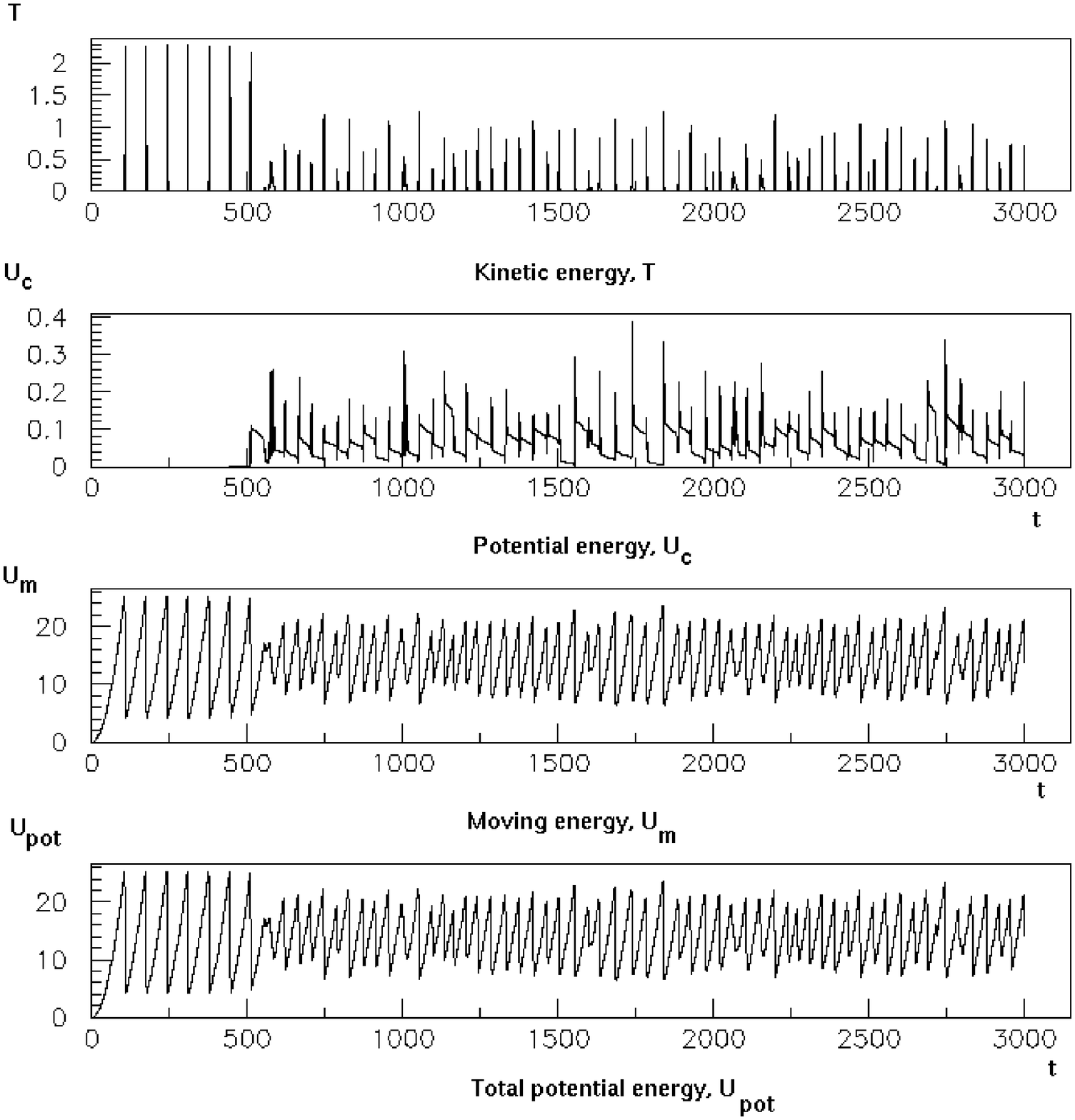,width=17cm}}}
\vspace{0.5cm}
\label{linener:pic}
\caption{Potential and kinetic energies for the linear 
         ($\epsilon =0$) N=100 block BK system
          $l=10, F_0=1, D=5,H=0.001, E=1$.}
\end{figure}
\begin{figure}[h]
\centerline{\mbox{\epsfig{file=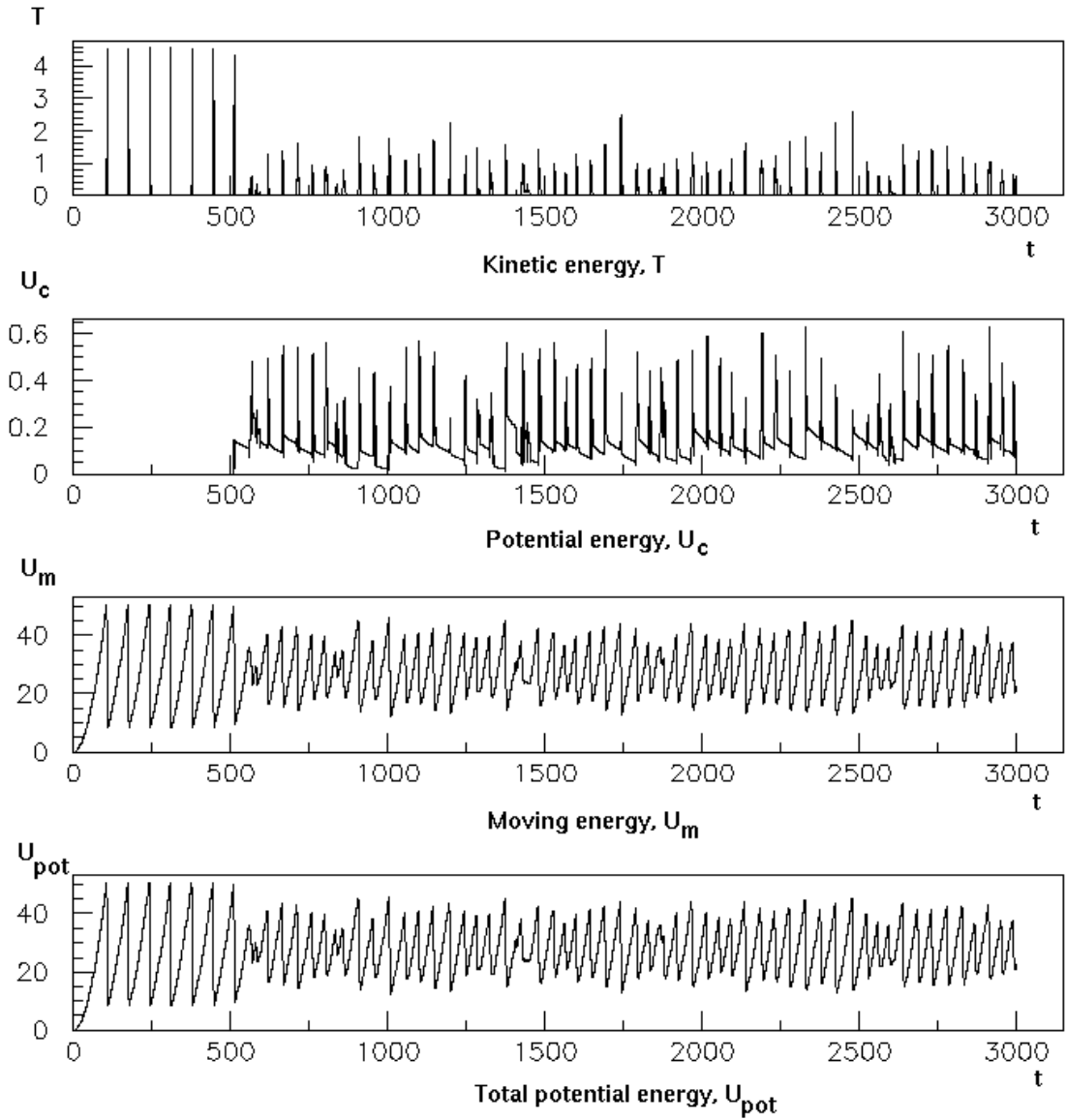,width=17cm}}}
\vspace{0.5cm}
\label{nelener:pic}
\caption{Potential and kinetic energies for the nonlinear 
         ($\epsilon =0.2$), N=100 block BK system
          $l=10, F_0=1, D=5,H=0.001, E=1$.}
\end{figure}
As it can be seen from the pictures, after the initial period of periodic 
slip-stick motion of the system as a whole (the typical period 
of one cycle is about loading time $\tau_L = 1/\nu$, $\tau_L=100$ 
for our simulations), the process becomes unstable and the system 
enters the chaotic regime.

As it was shown in the original paper of R.~Burridge and L.~Knopoff, 
the spring block system described by the equation (\ref{bk}) has 
a trivial solution $\forall n: u_n(t) = \nu t + F(\nu)$. This solution 
is unstable with respect to small perturbations. The time of instability 
of this uniform motion is comparable to the loading time $\tau_L$, after 
which the system exerts a slip. 

In our numerical investigations we have found another type of instability. 
After a 
period of time the quasi-periodic motion,
when the system slides and sticks as a whole, also becomes unstable
and actually chaotic when most part of inter-block springs are exited 
starts. The typical time of this instability is approximately the same 
for both linear and nonlinear models, see Figs.~3 and 4. At this stage 
($\tau < 500$) the inter-block springs $k_c$ have not accumulated 
enough energy and the dynamic of the system is determined by nonlinear 
friction force. Later, when nonlinear terms have got sufficient energy 
and the role of cubic and quartic terms in (\ref{bknldl}) becomes 
significant, the nonlinear model shows stronger clusterization of 
events than the linear one. This fact was traced on the 
autocorrelation function (see Fig.~5.),
\be
r(z) = \frac{\bra x(t)x(t+z)\ket}{\bra x^2(t)\ket}.
\ee
The averaging was taking over the available time series of energies 
$0\le \tau \le 2000$. The increasing of correlations at intermediate times 
($z < 3\tau_L$) is clearly observed for nonlinear model.
\begin{figure}[h]
\label{corfun:pic}
\centerline{\mbox{\epsfig{file=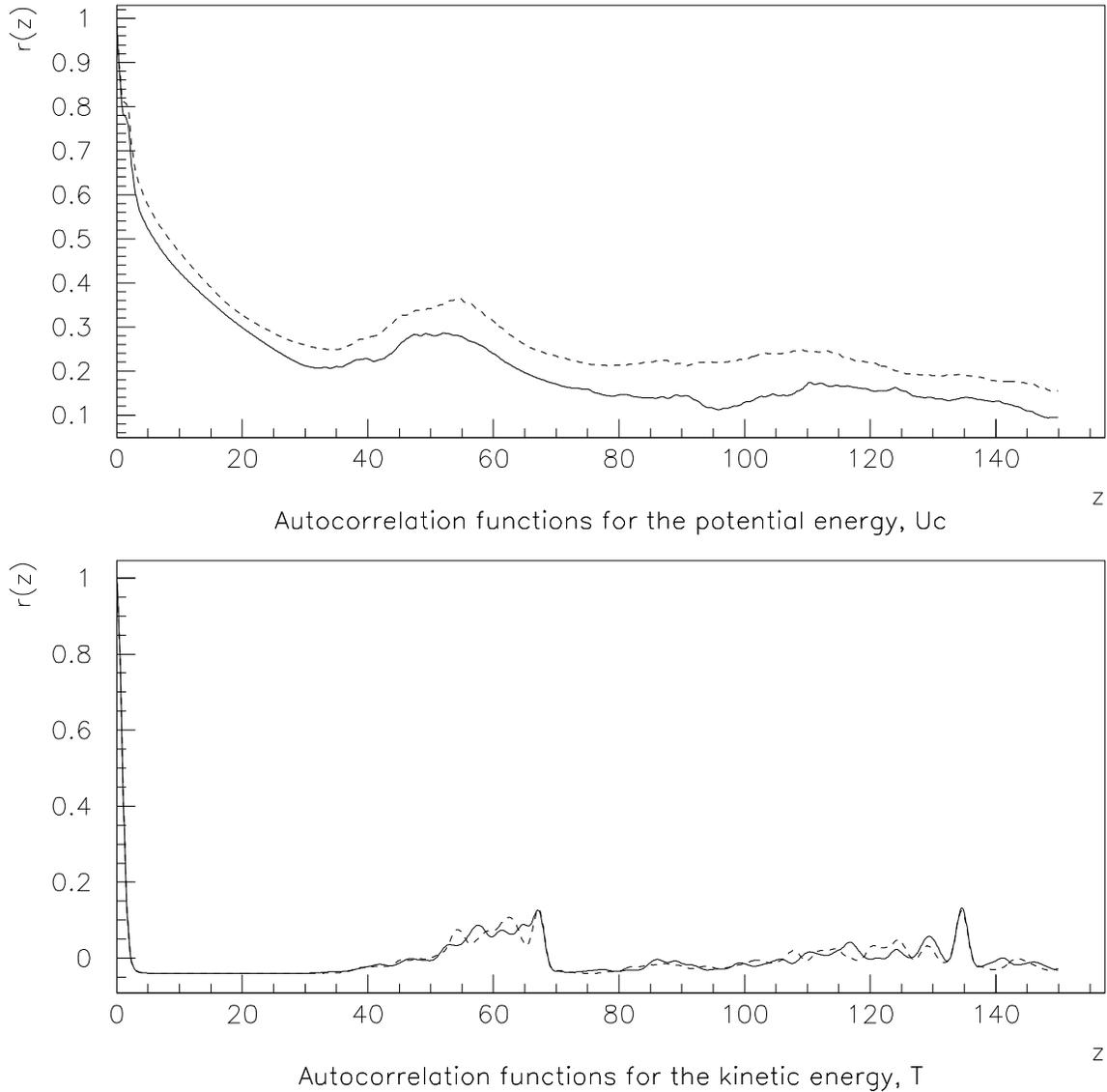,width=17cm}}}
\vspace{0.5cm}
\caption{Autocorrelation functions for potential and kinetic energies. 
         Shown for the nonlinear $\epsilon =0.1$,(dashed line) and 
         linear models. $N=10$ blocks for both curves}
\end{figure}

This leads to a collective behavior and probably provides an inverse 
energy cascade -- an energy flux from high frequency modes 
to low frequency ones, like that observed, say, in hydrodynamic turbulence. 
Besides that 
we can say the system to mimic {\em the liquification}
processes, taken place for real earthquakes. For strong earthquakes, 
long before the main earthquake event the foreshock motion 
of the crust are governed by elasticity equation; later -- in co-seismic 
stage -- we have more rheological behavior rather than elastic process.
The gravity waves like that on shallow water can be observed on the 
ground. The origin of these phenomena is that after 
a period of foreshocks the crust becomes less connected, less 
continuous and behaves as a system of particles with Van-der-Vaals 
interaction \cite{Lom94}. In this sense, the instability we have 
observed at $\tau$ about 1000 ($\nu =0.01, N=10$) is 
a solid-liquid phase transition, see Figs.~3,4.

\subsection{Clusterization}

The time dependence of potential and kinetic energies (\ref{pe},\ref{ke})
for linear ($\epsilon =0$) and  non-linear ($\epsilon =0.2$) model are 
presented on Figs.~3 and 4, respectively.

The nonlinear potential terms (\ref{U1}) contributes to 
``horizontal'' spring energies (second graph on both pictures). 
This potential terms, being relatively small in absolute values, 
play an essential role in nonlinear effects, happening in the 
spring-block chain. Initially, up to the time $t < 500$, the 
systems (both linear and non-linear) make a periodic slip-stick 
behavior as a whole. The potential energy of horizontal springs 
is very low at this stage: the internal degrees of freedom are not 
yet activated. Approximately at $\tau=500$ this periodic motion 
breaks down to quasi-periodic weakly chaotic regime. 

At this stage the nonlinear system shows stronger clusterization, 
than its linear counterpart; the groups of events looks more 
distinguished for nonlinear system, see Fig.~4. 

It can be seen from both potential energy of ``horizontal'' springs 
and the total potential energy (the graphs in the bottom). 
Possibly, for some other values of parameters a stronger intermittency 
will be observed. This, however, will be the subject of further 
investigation. Here we have just to note that nonlinear terms 
provide more coherence of seismic events (which can be observed 
as long wave modulation) and make the events stronger since more 
potential energy is accumulated.

\subsection{Solitons}

The dependence of displacement of coordinate and velocity as function
of time and number of blocks for the case of nonlinear 
model $\epsilon=0.1, N=10$ is presented on Fig.~6.
\begin{figure}[h]
\centerline{\mbox{\epsfig{file=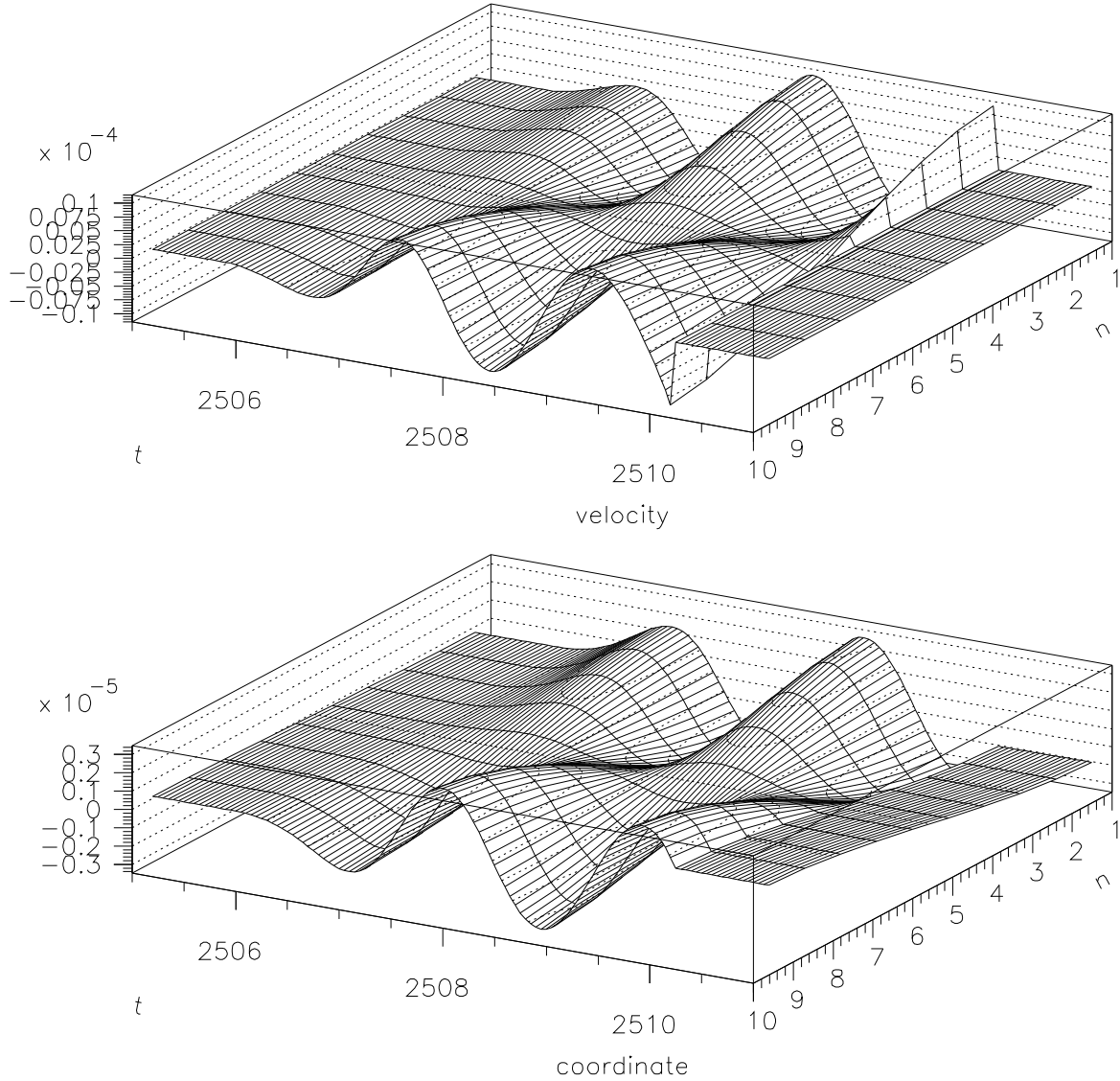,width=17cm}}}
\vspace{0.5cm}
\label{soliton:pic}
\caption{The displacements $u$ (lower picture) and velocities 
         $\dot u$ (upper picture) as functions of time and 
         block number. Calculated for nonlinear model of $N=10$ blocks, 
         $\epsilon=0.1$} 
         \end{figure}
The solitary waves moving from left to right 
are clearly seen on this pictures.

The localized solutions (propagating kinks 
$u(s,\tau) = u(\tau\pm s/\beta)$) of the BK system  
were already mentioned by Carlson and Langer as solutions of 
continuous limit equation of motion (formula 3.6 of \cite{CL89}):
$$(1-\xi^2/\beta^2)\ddot u = -u - \phi(2\zeta\nu+2\zeta\dot u).$$
In numerical simulations they were shown to coexist with chaotic 
modes. The properties of solitary wave solutions depend 
on sound velocity $\xi =al$, dissipation parameters, and crucially 
(as it was shown in \cite{SVR93} paper, specially devoted to 
solitons in BK model) on the dimensionless momentum parameter 
$\Theta = N\nu$, later generalized by the same authors \cite{SVR96} 
to the relation of loading time to traveling time $\theta$. 

Roughly, the number of solitary waves was numerically found to be $\Theta/8$. 
In our model there is a special source for solitary waves: 
the nonlinear spring terms, arising from self-similarity.

To study the effects of nonlinearity in the continuous 
limit of BK system, let us rewrite (1) in the 
co-moving frame $y_n = u_n -\nu\tau$:
\begin{eqnarray}
\nonumber
\ddot{y}_n &=& l^2(y_{n+1}-2y_n + y_{n-1}) -y_n  \\
           &+& \alpha\bigl[(y_{n+1}-y_n)^2-(y_n-y_{n-1})^2\bigr] \\
\nonumber  &+& \beta \bigl[(y_{n+1}-y_n)^3-(y_n-y_{n-1})^3\bigr] 
+F(2\zeta\dot{y_n} +2\zeta\nu).
\label{dls1}
\end{eqnarray}
Following Zabuzsky and Cruskal \cite{ZK65}, who studied solitons in
a similar system of differential equations, let us suppose that
at rest all the masses of relaxed system are located at points
$x_n = n a$, labeled by block number $n$, then we can use the Taylor
expansion for $y_{n+1}$ and $y_{n-1}$:
$$
y_{n\pm1}=y_n \pm a {y_n}' + \frac{a^2}{2!} {y_n}''
\pm \frac{a^3}{3!}{y_n}''' + \frac{a^4}{4!} {y_n}''''\pm \ldots.
$$
Substituting this into (\ref{dls1}) we obtain
\begin{equation}
\ddot{y}_n=l^2a^2{y_n}'' -y_n + 2\alpha a^3 {y_n}' {y_n}'' +
3\beta a^4 {{y_n}'}^2{y_n}'' + {a^4\over12}{y_n}''''
+\hbox{friction terms}.
\label{lateq}
\end{equation}

Up to the second and the forth terms at the r.~h.~s., the equation
(\ref{lateq}), we have driven at, coincides with the solitonic
equation of the Cruskal and Zabuzsky paper \cite{ZK65}. The latter, for
the case of small $\alpha$ can be transformed to Korteveg de Vries like 
equation
\begin{equation} \ddot{y}=y'' - 2\epsilon y' y'' +
{3\over2}\epsilon {y'}^2y'' + {a^2\over12}y'''' +\hbox{friction terms}..
\end{equation}
The detailed analysis of this equation will be done in further 
investigations.

However, we can stress either at this point that there is a 
principle difference between simple solitonic models of the  
paper  \cite{CL89} and our equation (\ref{lateq}). The former models come from 
no-scale approximation of BK model, which does not 
contain any characteristic scales. In this sense it much resembles the 
(Kolmogorov) universal regime of hydrodynamic turbulence\cite{K41a}. 
The equation (\ref{lateq}) comes from the finite grid approximation of 
an explicitly scale dependent (but self-similar) model, which is more 
adequate for seismic events with strong deformations.

\subsection{The distribution of events of different energies}

The original BK model was well tested to simulate empirical 
Gutenberg-Richter law \cite{GR44}, the relation between 
the probability of event and its seismic moment of the 
form 
$$\log N = a - b \log M, \quad \hbox{where}\quad b \approx 1.$$
In BK model simulations the seismic moment $M$ was understood 
as a sum of displacements taken over all blocks 
$M = \sum_i u_i$ \cite{CL89}.
In original paper \cite{BK67} the linear dependence 
between  the logarithm of cummulative number of events with 
seismic moment greater or equal to $M$ and $\log M$ holds only 
for the events of intermediate magnitude.
We have tested both linear and nonlinear model in the same way 
and get the same conclusion about the domail of its validity. 
The distribution of events of different magnitude is presented in 
Fig.~7.
\begin{figure}[h]
\centerline{\mbox{\epsfig{file=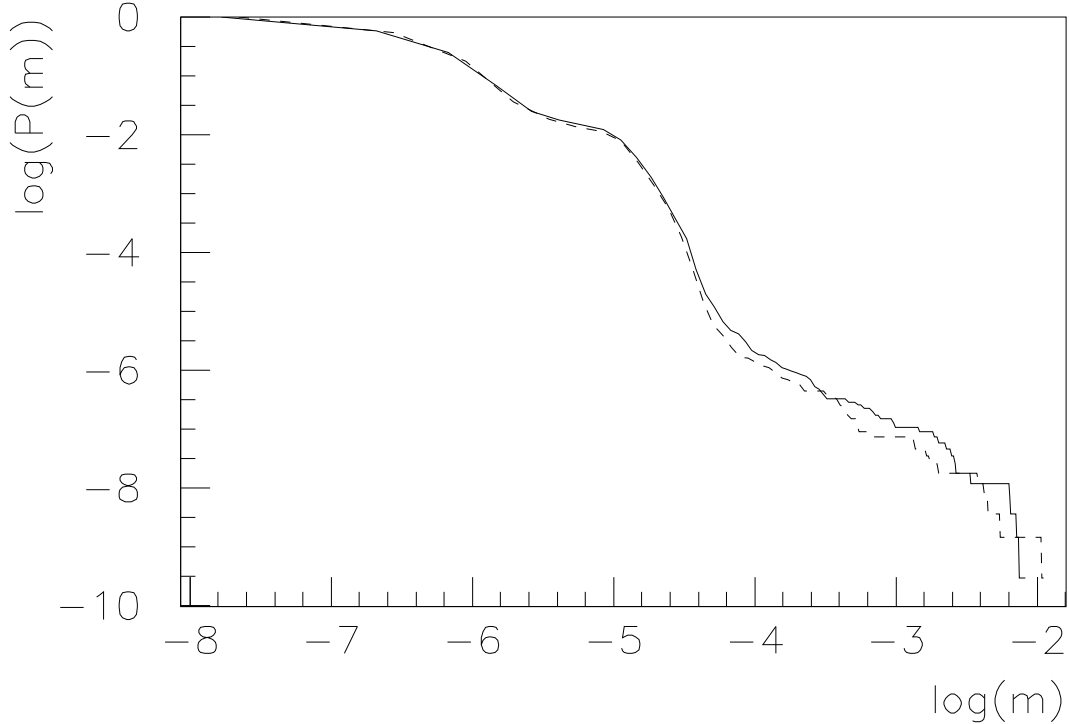,width=17cm}}}
\vspace{0.5cm}
\label{GR:pic}
\caption{The logarithm of normalized number of events with seismic 
         moment greater or equal to $M$ vs. $\log M$.
         Solid line is used for linear model, dashed line for nonlinear 
         model  of 10 blocks.} 
         \end{figure}

\subsection{Hurst exponents}

An important characteristics of any natural hazard process is the
Hurst exponent of its strength \cite{Feder,Lom94}. That means the time 
power-law behavior of the maximal deviation versus dispersion ratio
\be 
\frac{R(z)}{S(z)} = \left( {z\over2} \right)^h,
\label{RS}
\ee
where $R(z)$ is maximal deviation taken place for 
$0 < t \le z$.
$$R(z) = \max_{0< t \le z}X(t,z) - \min_{0<t\le z} X(t,z),$$ 
where 
$$X(t,z)|_{t<z} = \sum_{i=1}^z(\xi(i)-\langle \xi \rangle_z)$$
is the accumulated deviation 
for the same time period. In our model it will be accumulated 
potential energy. The square mean deviation for the process $\xi(t)$ in the 
time domain $0 < t \le z$ is 
$$S(z) = \sqrt{
          {1\over z}\sum_{m=1}^z (\xi(m)-
\langle \xi \rangle_z)^2
          }.$$

For Brownian motion, a purely random process, $h=1/2$. For other
stochastic processes, $H$ can be less or greater than $1/2$
(see \cite{Feder} for general explanation of this point).
The Hurst exponents for kinetic and potential energy of $N=10$ 
block ($\epsilon =0$) system calculated up to $z = 2000$ are 
presented in Fig.~8. The calculations have been done 
in a straightforward way by the definitions given above.
\begin{figure}[h]
\centerline{\mbox{\epsfig{file=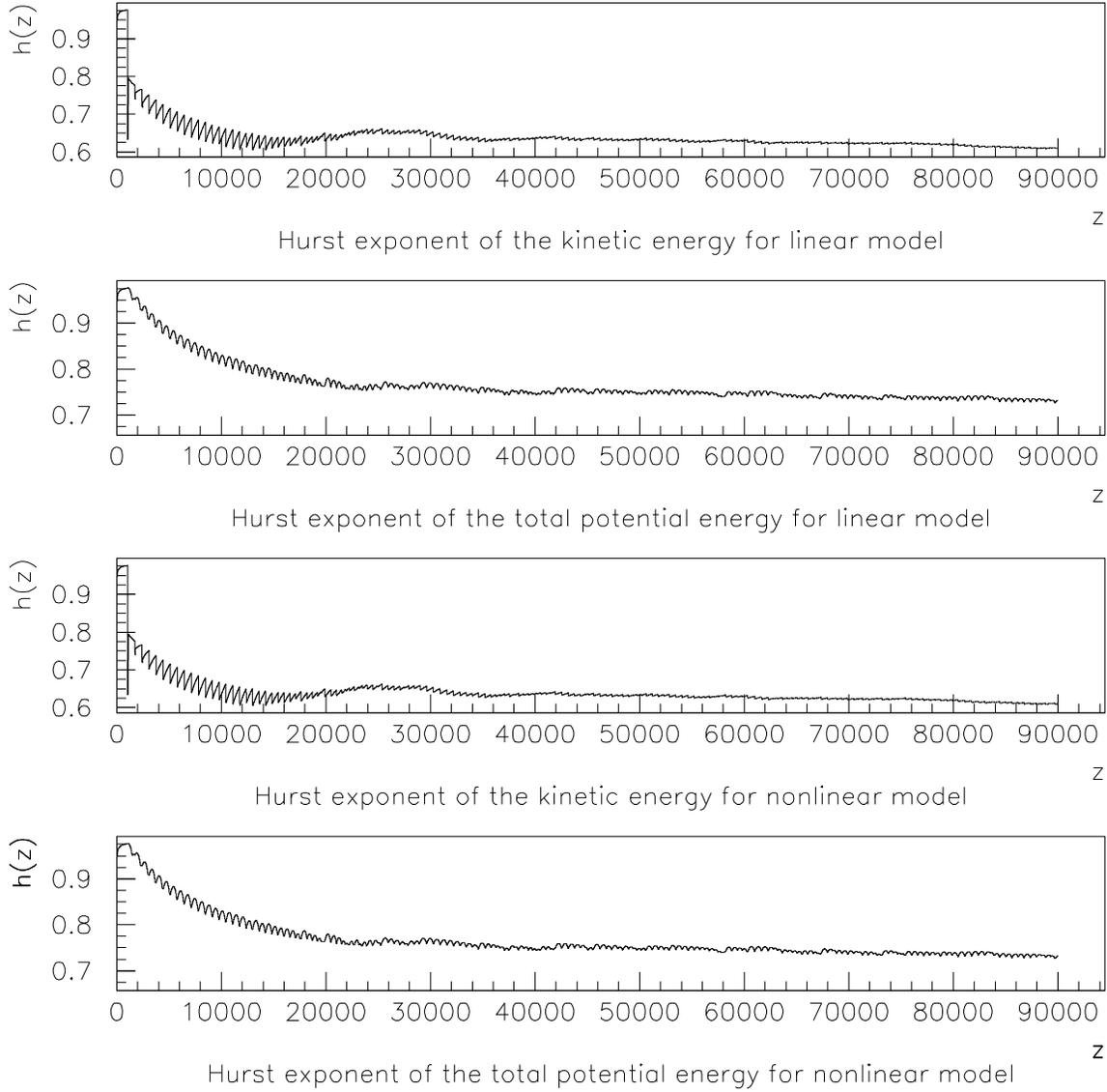,width=17cm}}}
\vspace{0.5cm}
\label{hurst:pic}
\caption{Hurst exponents for kinetic and potential energies. 
         Linear system of 10 blocks} 
\end{figure}
The Hurst exponents $h(z)$ may be found from the relations (\ref{RS}) 
for discrete time argument.
For both linear and nonlinear models
we can see from Fig.~8 that for the Burridge-Knopoff system 
$H$ exponent is much greater than 0.5. It lies about 0.6-0.8,
stabilizing for lower values about 0.7, when time increases.
This fact shows the persistence tendency of the simulating processes. 
This is very close to the value of Hurst exponent obtained 
for real earthquakes, which is also about 0.7 \cite{Lom94}.

\section{Dynamic description of BK system}

To study the dynamical properties of BK model, i.~e. the 
interaction of coupled blocks comprising the spring 
block chain, their effect on the motion of the system as a 
whole we rewrite the system (\ref{bk}) in 
vector form: 
\be
\ddot{\bf u} = -l^2 [A] {\bf u} - ({\bf u} - \nu t{\bf e}_0) + {\bf F}(\dot {\bf u}),
\label{vbk}
\ee
where $[A]$ is $[N\times N]$ matrix 
$$[A] = \pmatrix{ 1 & -1 & 0 & 0  & 0 & \cdots & 0 \cr
                    -1 & 2  &-1 & 0  & 0 & \cdots & 0 \cr
                     0 & -1 & 2 & -1 & 0 & \cdots & 0 \cr
    \cdots & \cdots & \cdots & \cdots & \cdots & \cdots & \cdots \cr
                      0 & \cdots & 0  & -1& 2  & -1& 0  \cr
                      0 & \cdots & 0  & 0 & -1 & 2 & -1 \cr
                      0 & \cdots & 0  & 0 & 0 & -1 & 1 }
$$
and
$$
{\bf u} = (u_1,\ldots,u_N)^T,\quad {\bf F} = (F({\dot u}_1),\ldots,F({\dot u}_N))^T,\quad {\bf e}_0 = (1,1,\ldots,1)^T.
$$
Let $\{ \lambda_i, {\bf e}_i \}$ be the eigenvalues and eigenvectors of $[A]$:
\be 
[A] {\bf e}_i =  \lambda_i {\bf e}_i,\quad i = \overline{1,N},\quad 
({\bf e}_i,{\bf e}_k) = \delta_{ik},
\label{es}
\ee
where
$ \lambda_1\le \lambda_2 \le \ldots \le \lambda_n$.

Using the representation ${\bf u} = \sum_{i=1}^{N} s_i{\bf e}_i$ we rewrite 
the equations of motion of Burridge-Knopoff model (\ref{vbk}) as: 
\be 
\ddot{\bf s}_i = -(\lambda_i l^2+1){\bf s}_i + ({\bf e}_0,{\bf e}_i) \nu t 
+ ({\bf e}_i,{\bf F}), \quad  i = \overline{1,N},
\ee
where $s_i(t) = ({\bf u}(t),{\bf e}_i)$.
It should be noted that matrix $[A]$ has exactly $N$ different nonnegative 
eigenvalues 
$ \lambda_1 < \lambda_2 < \ldots < \lambda_N$ with $\lambda_1=0$. The eigenvector 
corresponding to the lowest eigenvalue is collinear to ${\bf e}_0$:
$$ 
{\bf e}_1 = \frac{1}{\sqrt N}{\bf e}_0.
$$ 

Thus, we reduce the original problem to the system of nonlinear oscillators interlinked by nonlinear friction force ${\bf F}$. Using Green functions for
one-dimensional Helmgholtz equation
$\ddot{u}+\omega^2 u = 0$
and the orthogonality 
of basic vectors $\{{\bf e}_i\}_1^N$ we express the solution of BK system (\ref{vbk}) in 
the form: 
\be
{\bf u}(t) = \nu t {\bf e}_0 + \sum_{i=1}^N {\bf e}_i \Bigl[
 a_i\sin \omega_i t + b_i\cos\omega_i t
     + \int_0^t \frac{\sin(\omega_i(t-x))}{\omega_i}({\bf e}_i,{\bf F}(\dot{\bf u}(x))) dx
                                                      \Bigr],
\label{gf1}
\ee
where $\omega_i^2 = \lambda_il^2+1$. Constants $a_i$ and $b_i$ depend on 
the initial state of the BK model.

As one can see from this equation, the behavior of the solution ${\bf u}(t)$ 
is governed by projections of dissipative force 
${\bf F}$ onto the basis  $\{{\bf e}_i\}_1^N$. 
In Figs.~9,10 the time dependence of projections 
$({\bf e}_i,{\bf F}(\dot {\bf u}(x))), i=\overline{1,3}$ are shown 
for the case of linear model, $\epsilon =0$.
\begin{figure}[h]
\centerline{\mbox{\epsfig{file=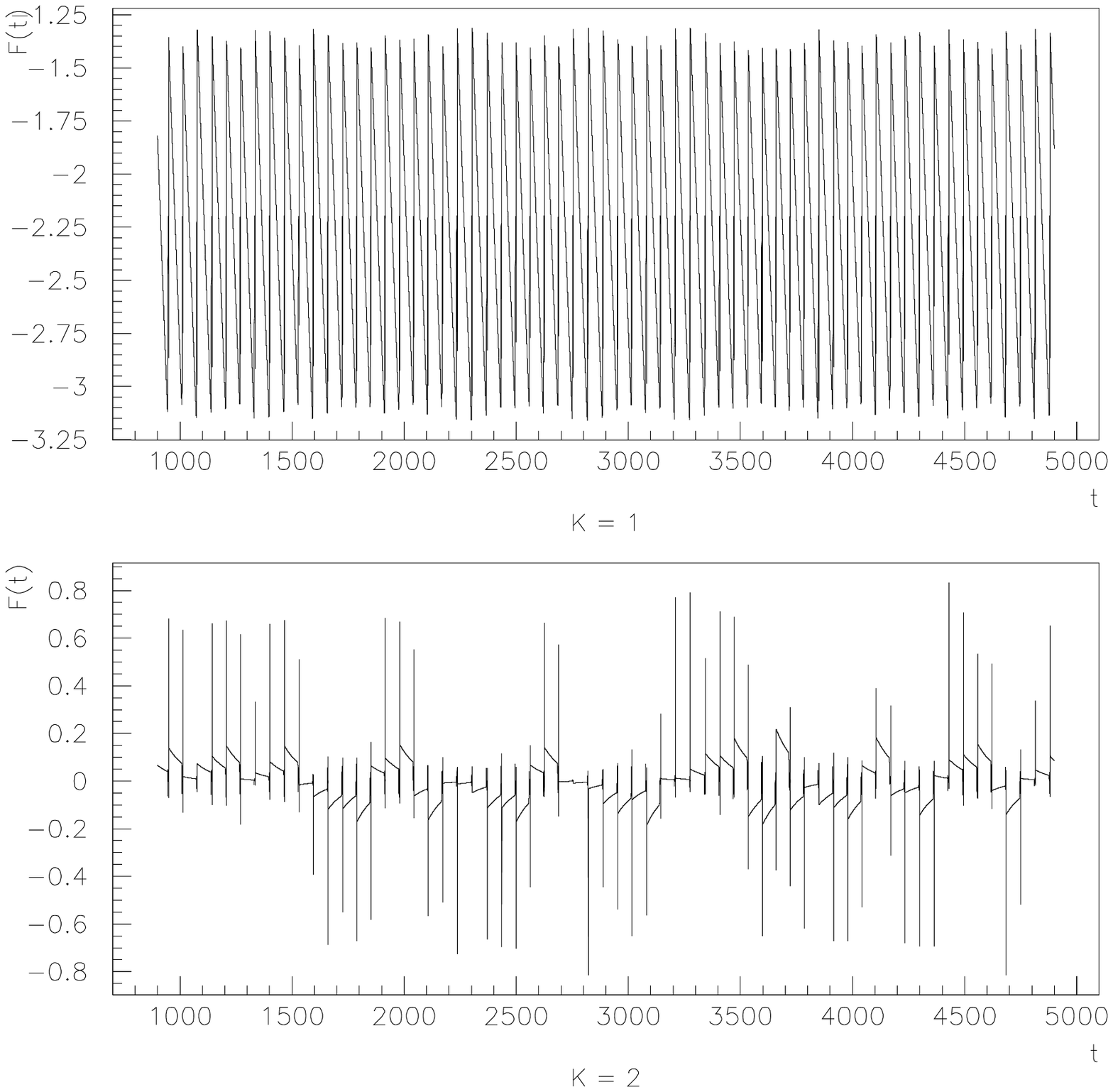,width=17cm}}}
\vspace{0.5cm}
\label{fc1:pic}
\caption{The projections of dissipative force on first two eigenvectors 
         of the system (\ref{es}). A 10 block system.} 
\end{figure}
\begin{figure}[h]
\centerline{\mbox{\epsfig{file=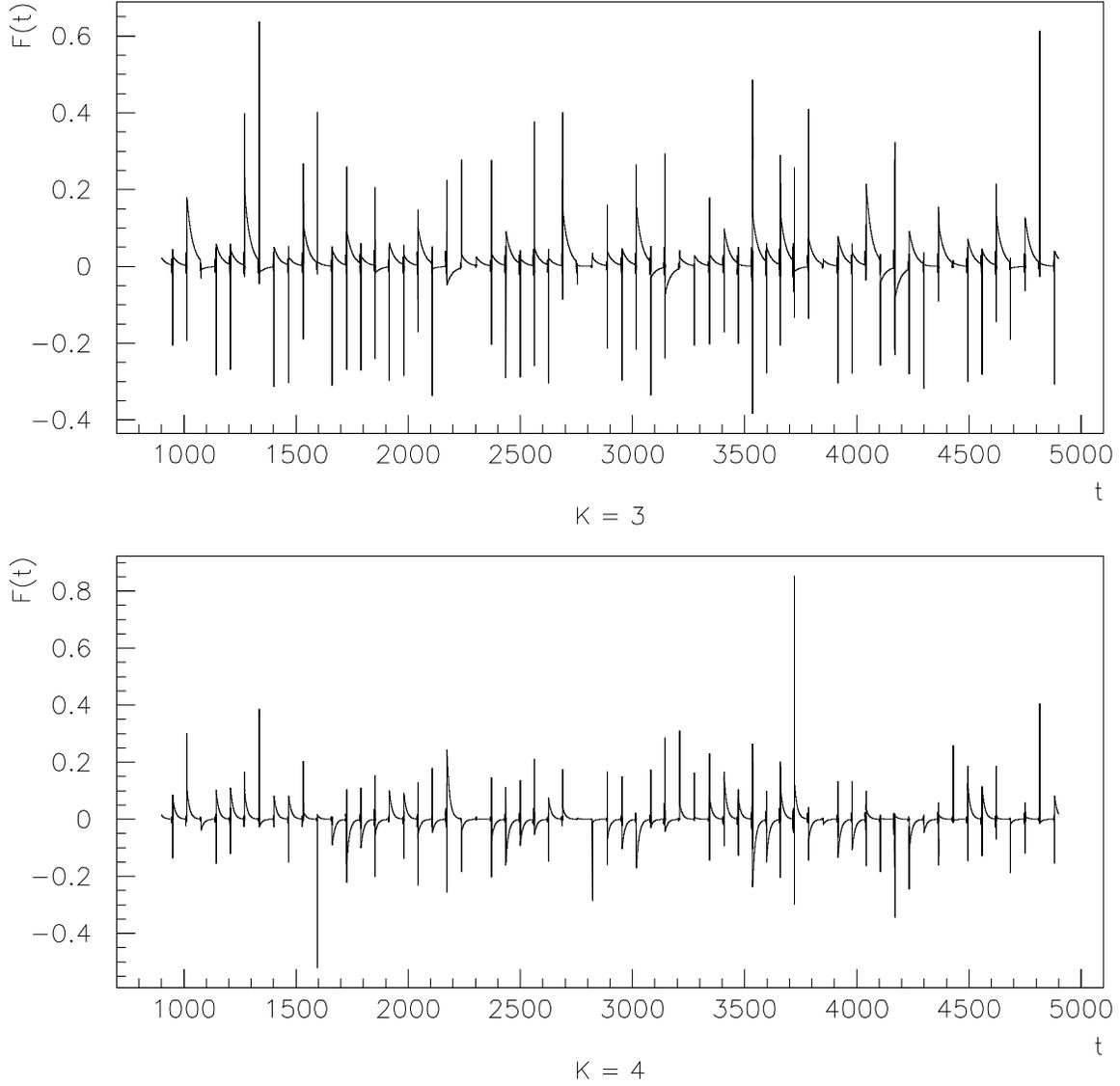,width=17cm}}}
\vspace{0.5cm}
\label{fc2:pic}
\caption{The projections of dissipative force on 3d and 4th eigenvectors 
         of the system (\ref{es}). A 10 block system.} 
\end{figure}
The projection of the first eigenvector looks quasi-periodic, all others exhibit chaotic behavior, like a randomly kicked oscillator \cite{Zasl70}. 

For the first harmonics the convolution in r.~h.~s.\, of (\ref{gf1}) is difficult to evaluate, and we use 
another representation to study its behavior qualitatively. Namely, we use the Green function for the equation  $\ddot s = g(t)$.
Doing so, we obtain
\be
\begin{array}{lcl}
{\bf u}(t) &=& \nu t {\bf e}_0 + \sum_{i=2}^N {\bf e}_i \Bigl[ a_i\sin \omega_i t + b_i\cos\omega_i t 
+ \int_0^t \frac{\sin(\omega_i(t-x))}{\omega_i} ({\bf e}_i,{\bf F}(\dot {\bf u}(x)) dx\Bigr] \\
           &+& {\bf e}_1\Bigl(a_1 + b_1 t + ({\bf e}_1,\int_0^t (t-x)({\bf F}(\dot{\bf u}(x)-{\bf u}(x) + \nu x {\bf e}_0) dx) \Bigr).
\end{array}
\label{gf2}
\ee  
 The function $G(t)=(({\bf F}(\dot{\bf u}) - {\bf u} +
\nu t {\bf e}_0), {\bf e}_1)$ is shown in Fig.~11. 
This function looks like the above 
considered projections $({\bf F},{\bf e}_2)$, $({\bf F},{\bf e}_3)$, etc.
\begin{figure}[h]
\centerline{\mbox{\epsfig{file=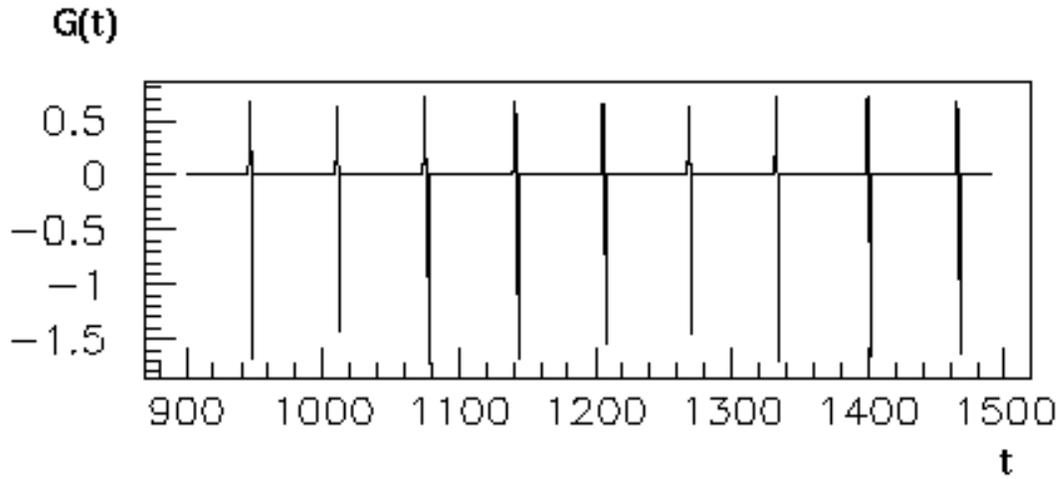,width=17cm}}}
\vspace{0.5cm}
\label{Gt:pic}
\caption{Time dependence of driving term G(t).}
\end{figure}

The phase portrait $(S_1, {\dot S}_1)$ of the first harmonics $s_1$ is presented in co-moving frame 
$$
  S_1(t) = s_1(t) - \nu t ({\bf e}_0,{\bf e}_1)
$$
in Figs.12,13 at different times for linear ($\epsilon = 0$) and nonlinear 
($\epsilon = 0.1$) models, respectively.
At the absence of friction force all orbits are evidently that of harmonic oscillator, i.~e. ellipses. 
It should be noted that the motion of first mode $S_1$ up to a constant multiplier coincides with 
the motion of center of masses in co-moving $\nu t$ frame.  
\begin{figure}[h]
\centerline{\mbox{\epsfig{file=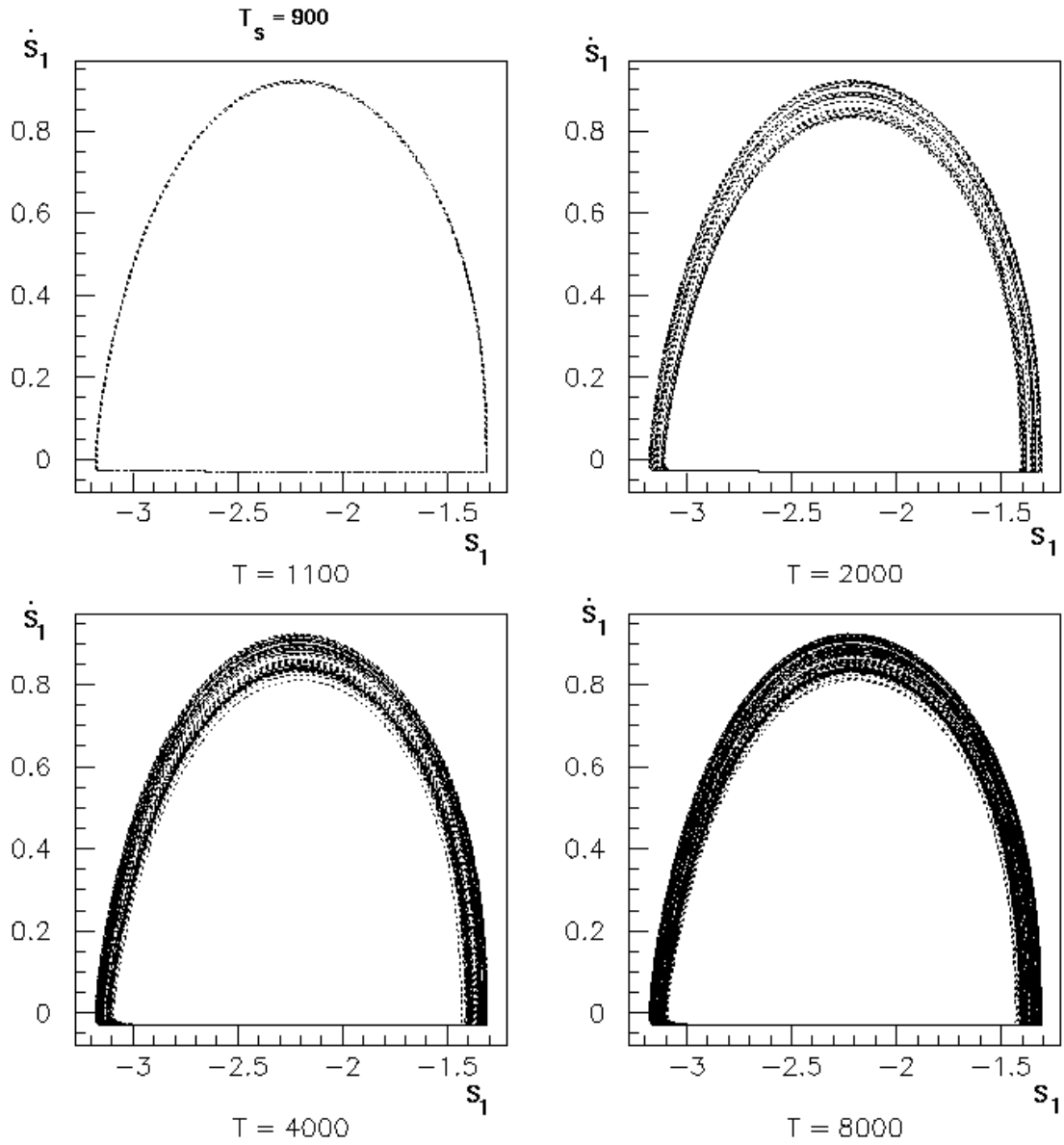,width=17cm}}}
\vspace{0.5cm}
\label{fas1l:pic}
\caption{The phase portrait  $(S_1, {\dot S}_1)$ for linear system of 
10 blocks.}
 \end{figure}
\begin{figure}[h]
\centerline{\mbox{\epsfig{file=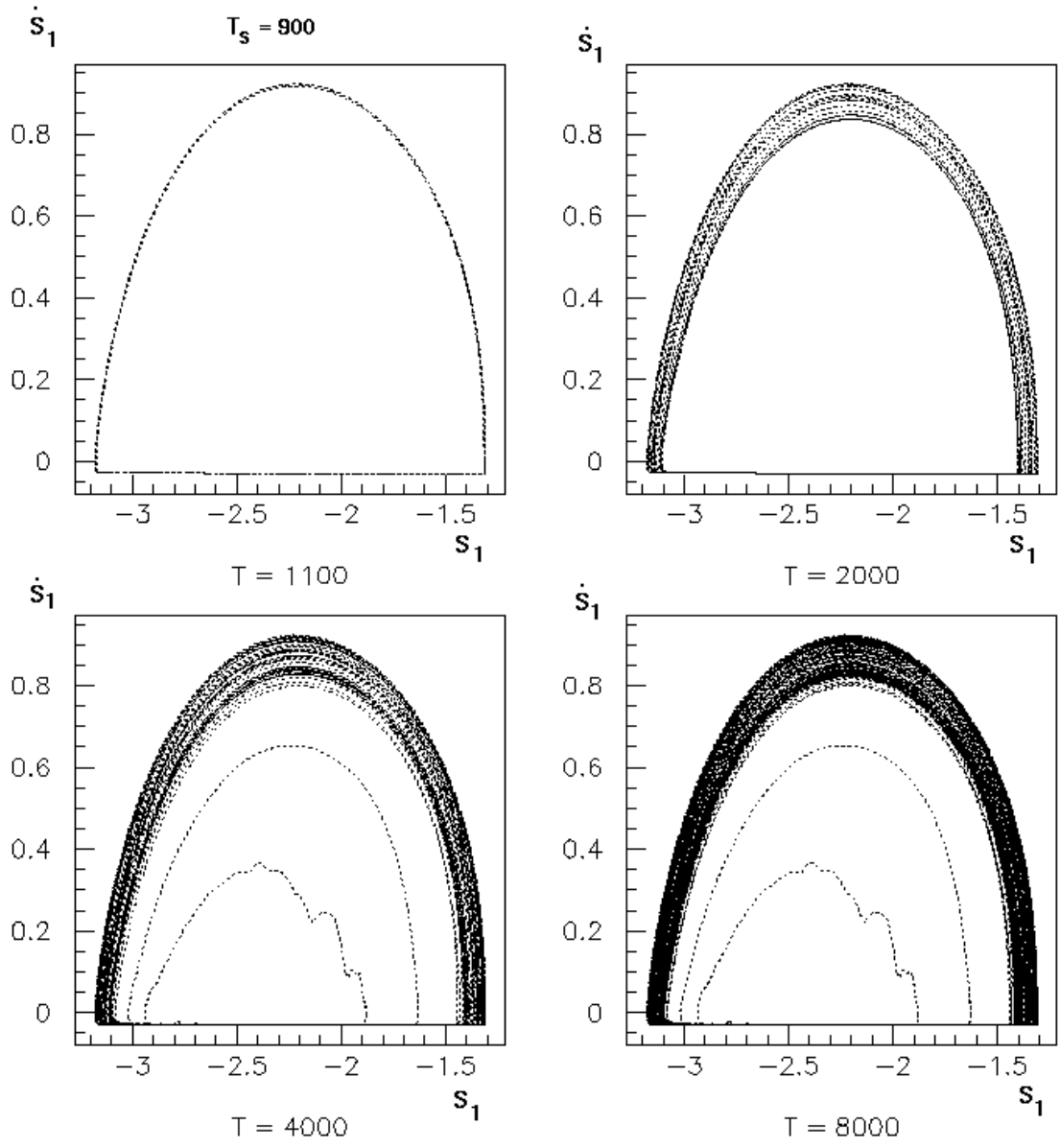,width=17cm}}}
\vspace{0.5cm}
\label{fas1n:pic}
\caption{The phase portrait  $(S_1, {\dot S}_1)$ for nonlinear system of 
10 blocks.}
 \end{figure}

At sufficiently large times, above $t\,>\,3000$, the nonlinear system 
departs away from quasi-periodic attracting set of the linear model 
(compare Fig.~12 and Fig.~13). The horse-shoe like cycles shown in 
Figs.~12,13,14 and 
15 evidently have self-similar structure if zoomed at different scales. 
The next-to-first harmonics ($s_2$) phase portrait of nonlinear model 
(Fig.~15) is, in contrast, compressed in comparison to its linear 
counterpart Fig.~14.
\begin{figure}[h]
\centerline{\mbox{\epsfig{file=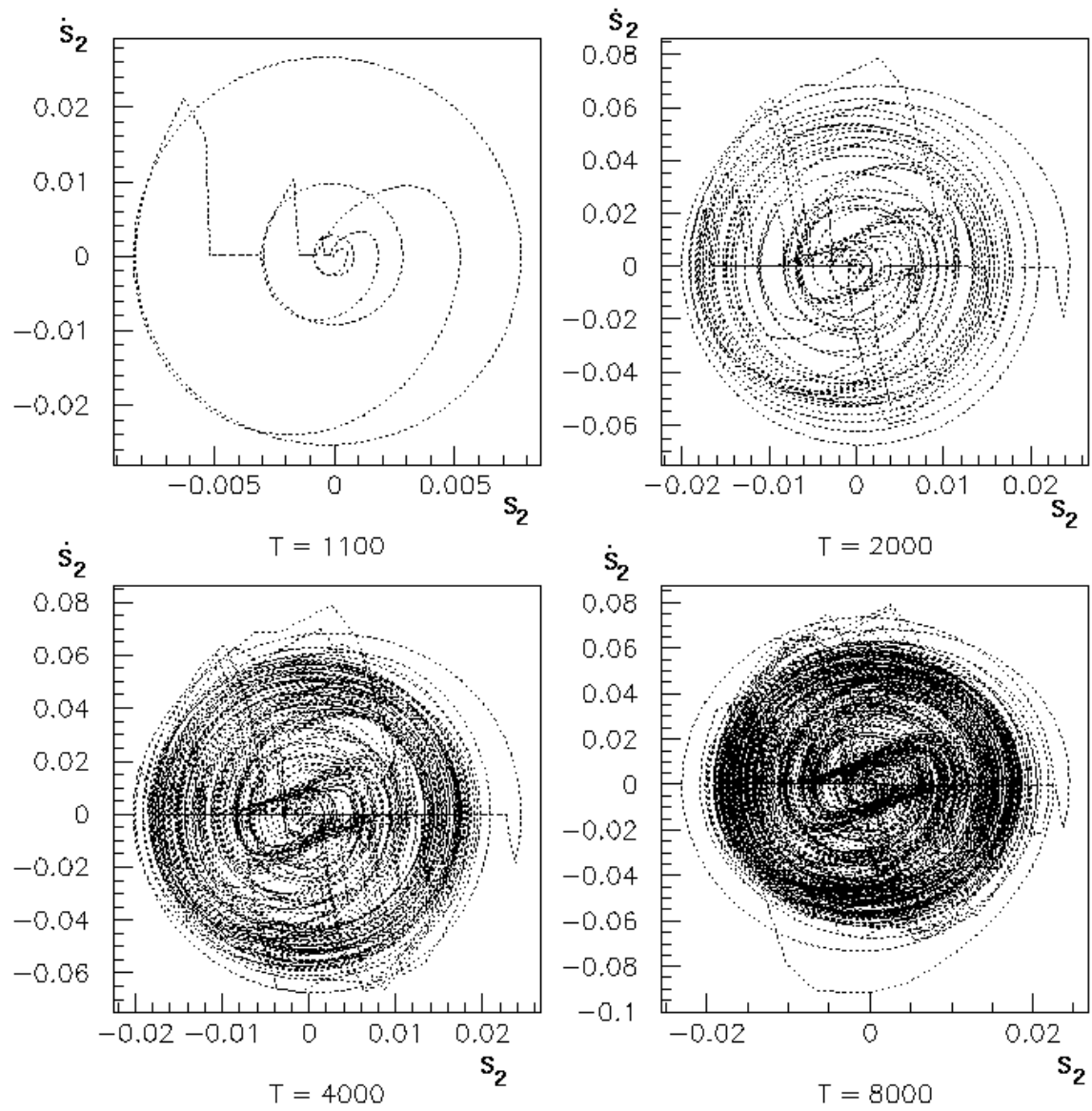,width=17cm}}}
\vspace{0.5cm}
\caption{The phase portrait  $(S_2, {\dot S}_2)$ for linear system of 10 
blocks.}
\label{fas2l:pic}
 \end{figure}
\begin{figure}[h]
\centerline{\mbox{\epsfig{file=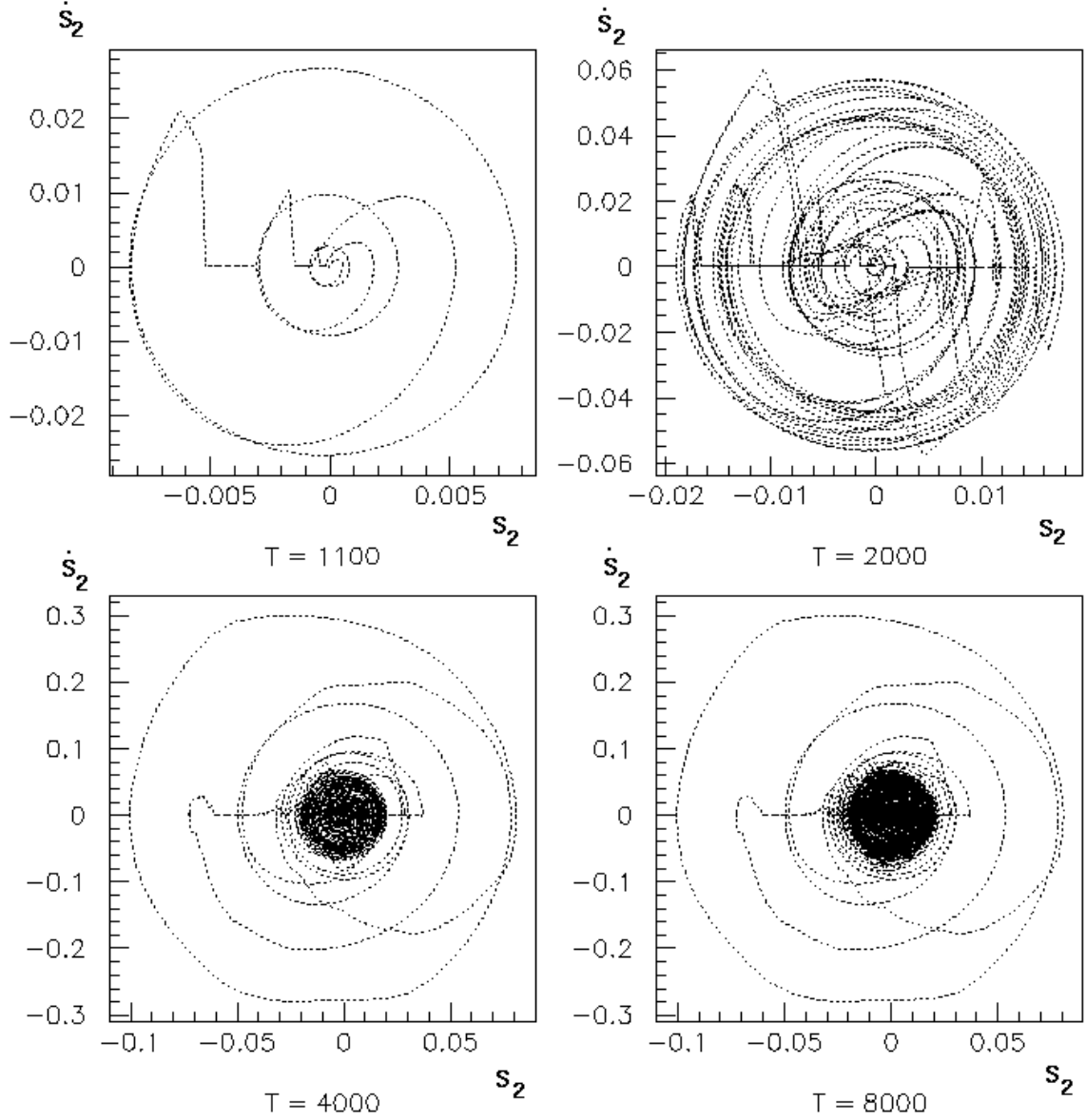,width=17cm}}}
\vspace{0.5cm}
\label{fas2n:pic}
\caption{The phase portrait  $(S_2, {\dot S}_2)$ for nonlinear system  of 
10 blocks.}
\end{figure}  
The higher modes $s_k, k>2$, have the same behavior.  

Therefore, we infer that the nonlinearity causes the energy transfer from 
individual block oscillations to the large-scale collective motions. 
This very much resembles gravity waves observed in co-seismic stage of 
strong earthquakes \cite{Lom94}. These waves, looking like shallow 
water gravity waves are often developed on soft crust, where the 
nonlinearity is much higher than on rigid, i.e. rocky crust.

The representation of the solution $u(t)$ in the form 
$u(t)=\sum _{i=1}^{N} s_i(t) \vec {u_i}$ allows to describe separately 
the behavior 
of the BK model as a whole and the interaction of blocks. This approach 
may be useful in futher studies devoted to the BK model.

\section*{Conclusion}
The dynamical behavior of BK mechanical model of earthquake 
faults was investigated. In contrast to standard BK we incorporate 
nonlinear terms in inter-block springs potential to account for 
self-similarity, which is widely observed for earth crust elastic 
properties. We have shown that nonlinearity arising 
from self-similarity can be considered as an additional source 
of solitonic behavior such as gravity waves observed in co-seismic
stage of strong earthquakes. We have observed numerically that 
quasi-periodic slip-stick motion of the spring-block chain as a whole
after a period of about $10^2\times\tau_L$ breaks to chaotic 
behavior. This is much like a liquification process which plays 
an important role for earthquake caused disasters on soft ground. 
The phase analysis of the model shows sinchronization of phases 
of different blocks, which causes strong coherent motions of 
the system as a whole -- large seismic events.  

So, being very simplistic the Burridge-Knopoff model imitates 
chaotization, related to the decay of low frequency modes 
to a number of high frequency ones, and the inverse processes --
the formation of low frequency modes from differences of high 
frequency harmonics -- known as inverse energy cascade.

\section*{Acknowledgements}
The authors are grateful to Dr. G.~M.~Molchan for useful discussions.
This work was supported by the European Commision.

\end{document}